\documentstyle[preprint,prl,aps]{revtex}

\begin{document}

\draft

\title{Nonmonotonous magnetic field dependence and scaling of the
thermal conductivity for
superconductors with nodes of the order parameter}

\author{Yu.~S.~Barash $^{a^{\dag},b^{\ddag}}$ and
A.~A.~Svidzinsky $^{a,c^{\ddag}}$
\\}
\address{
$a)$ P.N. Lebedev Physical Institute, Leninsky Prospect 53, Moscow 117924,
Russia \\
$b)$ DRFMC, Service de Physique Statistique, Magn\'etisme et
Supraconductivit\'e, CENG, \\
17, rue des Martyrs, 38054 GRENOBLE - Cedex 9, France\\
$c)$ Department of Physics, Stanford University, Stanford CA 94305--4060, USA\\
}
\date{\today}

\maketitle

\begin{abstract}
We show that there is a new mechanism for nonmonotonous behavior of magnetic
field dependence of the electronic thermal conductivity $\kappa$ of clean
superconductors with nodes of the order parameter on the Fermi surface. In
particular, for unitary scatterers the nonmonotony of relaxation time takes
place. Contribution from the intervortex space turns out to be essential for
this effect even at low temperatures. Our results are in a qualitative
agreement with recent experimental data for superconducting
$UPt_3$\cite{fl1,fl2}. For $E_{2u}$-type of pairing we find approximately the
scaling of the thermal conductivity in clean limit with a single parameter
$x=\frac{T}{T_c}\sqrt{\frac{B_{c2}}{B}}$ at low fields and low temperatures, as
well as weak low-temperature dependence of the anisotropy ratio
$\kappa_{zz}/\kappa_{yy}$ in zero field. For $E_{1g}$-type of pairing
deviations from the scaling are more noticeable and the anisotropy ratio is
essentially temperature dependent.
\end{abstract}


\newpage

\section{Introduction}

Growing amount of experimental data indicates that many of high-temperature and
heavy-fermion superconductors have highly anisotropic order parameter on the
Fermi surface.  The possibility  for the order parameter to have opposite
signs in different regions on the Fermi surface, lines of nodes between those
regions (and, possibly, point nodes too) has attracted much attention. Since
the presence of nodes of the order parameter leads to the existence of
low-energy excitations in spectrum, this strongly modifies low-temperature
behavior of thermodynamic and transport characteristics both in the absence and
under the applied magnetic field (~see reviews\cite{sigueda,sauls} and for more
recent literature, for example,\cite{vol1,vol3,kop1,bar1}~). For these reasons
measurements of temperature and magnetic field dependences of those
characteristics are important experimental tools for probe the anisotropic
structure of the order parameter\cite{fl1,fl2,mol,fgr,lus1,lus2,sud}.  As
compared to thermodynamic characteristics like specific heat, an important
advantage of studying transport properties, in particular, the thermal
conductivity is that it is a directional probe, sensitive to relative
orientations of the thermal flow, the magnetic field and directions to nodes of
the order parameter.

For isotropic $s$-wave superconductors low temperatures are obviously defined
as satisfying the condition $T\ll  \Delta_{max}$, when the number of
quasiparticles thermally activated above the gap is exponentially small.  In
the presence of nodes of the order parameter on the Fermi surface an additional
small energy scale $\gamma$ appears, describing the bandwidth of
impurity-induced quasiparticle bound states. Then the low-temperature region
should be divided into two parts. For temperatures $T\lesssim\gamma$ transport
properties are dominated by bound states. Non-zero density of these states on
the Fermi surface results in linear temperature dependence of the thermal
conductivity in this case. At the same time under the condition $\gamma\lesssim
T\ll \Delta_{max}$, which can be satisfied for sufficiently clean
superconductors, one can disregard the influence of bound states. Then the
presence of nodes of the order parameter leads to characteristic higher order
power-law behaviors of the thermal conductivity with
temperature\cite{hir1,hir2,graf1,graf2,nor1,pet1,pet2,bar2}.

The thermal conductivity of $s$-wave isotropic superconductors in the presence
of the applied magnetic field exhibits nonmonotonous dependence upon the
field\cite{saint,lowell,vinen,chak}. In high fields the thermal conductivity
$\kappa(B)$ rises rapidly due to suppression of the order parameter as the
magnetic field approaches $B_{c2}$.  At low fields $\kappa(B)$ decreases with
increasing magnetic field.  This phenomenon is usually attributed to the
scattering of electrons (or/and phonons) by the fluxoids, which was
theoretically considered in\cite{cleary}. However, in isotropic $s$-wave
superconductors at low temperatures the number of quasiparticles thermally
activated into scattering states are exponentially small, so that the
contribution to the thermal conduction along the magnetic field from bound
states within the vortex cores become essential. In contrast to isotropic
superconductors, for anisotropically paired superconductors with nodes of the
order parameter on the Fermi surface, the intervortex space can dominate
thermodynamic and transport characteristics at low temperatures due to low
energy excitations with momentum directions near the
nodes\cite{vol3,kop1,kop2,plee,kop3}. Under these conditions the
characteristics usually become quite sensitive to the applied magnetic field
even if it is of relatively small value. For example, quasiparticle density of
states on the Fermi surface for clean superconductors with nodes, takes nonzero
value just due to the applied magnetic field (~as well as impurities~).  Since
low-temperature behavior of thermodynamic quantities like the specific heat is
directly associated with the behavior of the density of states on the Fermi
surface $N_s(0)$, this is the reason for their sensitivity to the magnetic
field\cite{vol1,vol3}. For transport characteristics additional quantities are
of importance in this respect. These are scattering relaxation times for
various channels. For instance, it is known that the relaxation time
$\tau_s(\omega)$ at sufficiently low energies for elastic impurity scattering
in clean superconductors with nodes, may be quite small for unitary scatterers
and extremely large in the case of Born scatterers\cite{rice,pet3,pet1,pet2}.
While $\tau_s(\omega)$ for Born scatterers is directly associated with
$N_s(\omega)$, in the unitary limit $\tau_s(\omega)$, as a function of energy,
is not reduced entirely to the density of states and should be considered as an
independent quantity.

Below we show that for anisotropically paired clean superconductors with nodes
of the order parameter on the Fermi surface, at low temperatures there is
important additional mechanism for nonmonotonous dependence of the electronic
thermal conductivity upon the magnetic field. This mechanism is associated with
the electronic contribution to $\kappa(B)$ mostly from the intervortex space
due to the presence there of extended quasiparticle states of low energies with
momentum directions near nodes of the order parameter.  Contrary to isotropic
$s$-wave superconductors, the main effect comes in this case from the influence
of condensate flow field (even of relatively small value) on the quasiparticle
impurity scattering in the unitary limit, rather than from scattering of
quasiparticles directly on vortex cores.  For these two types of scattering
inverse relaxation times could be added under certain conditions\cite{cleary}.
However, the contribution from scattering of quasiparticles by vortex cores is
supposed to be negligibly small under the conditions considered below, as
compared to the one from scattering by impurities in the intervortex space.

In the unitary limit relaxation time for scattering by nonmagnetic impurities
of low-energy quasiparticles, is found below to be a nonmonotonous function
upon the condensate flow field. This takes place even for uniform superfluid
flow, that is in the absence of any scattering by vortices.  Furthermore, for
type II superconductors with large Ginzburg-Landau parameter, the superflow
induced by magnetic field in the intervortex space can be considered on
sufficiently large distances from vortex cores (much greater than the coherence
length) as quasihomogeneous flow.  This allows one to consider approximately
the thermal conductivity as a function of the local value of condensate flow
field, as this would be for the uniform flow.  Magnetic field dependence of the
thermal conductivity obtained in the first approximation as spatial averaging
of the result over the intervortex space, is found below to be nonmonotonous
under certain conditions.

This effect can be important, in particular, for the analysis of recently
observed nonmonotonous behavior of the thermal conductivity in $UPt_3$, since
the electronic contribution to $\kappa$ is known to dominate there below
1K\cite{fl1,fl2}. We consider $(1,i)$ phases both for the $E_{2u}$
representation and for the $E_{1g}$ one as candidates for the type of
superconducting pairing in $UPt_3$ at low temperatures and under the weak
applied magnetic field $B_{c1}< B\ll B_{c2}$ (see, for
example,\cite{sigueda,sauls,joynt}). Our theoretical results, basing on this
consideration, are in a qualitative agreement with those experimentally
observed in\cite{fl1,fl2}. Mostly the both models give rise to such an
agreement. The difference between corresponding predictions is not too great,
although it is of importance permitting to distinguish between them.  Recent
experimental data\cite{fl1,fl2}, in particular, allow for determining the
low-temperature interval, where the power law temperature dependence of the
thermal conductivity takes place. We find, that the behavior of the zero-field
anisotropy ratio for the thermal conductivity for these temperatures seems to
indicate in favor of $E_{2u}$ type of pairing.  At higher temperatures the
behavior of the thermal conductivity becomes essentially depending upon the
particular form of the order parameter all over the Fermi surface, not only
near its nodes.  Under this condition there are various possibilities to fit
experimental data within the framework of both models, so that the problem to
distinguish between them becomes ambiguous one\cite{nor1}.  By contrast, at
sufficiently low temperatures $T\ll T_c$ the behavior of the thermal
conductivity is governed mainly by the behavior of the order parameter near
nodes, as well as by the strength of scatterers. This leads, in principle, to
the possibility to identify the behavior near the nodes and, hence, the type of
superconducting pairing, but not a particular form of the order parameter all
over the Fermi surface. This circumstance was already emphasized earlier
in\cite{graf2}, where the accent was made on the ultra low temperatures
$T\ll\gamma$. Quite a small value $\gamma\approx 0.017K$ taking place for clean
samples of Ref.\ \onlinecite{fl1,fl2} give the possibility to determine and
concentrate on the low-temperature region $\gamma\lesssim T\ll T_c$, while the
ultra low temperatures in this case seem not to be sufficiently studied
experimentally yet.

The article is organized as follows. Basic equations are listed in the next
section. We consider clean superconductors at low temperatures under the
applied magnetic field (~$H_{c1}\lesssim B\ll H_{c2}$~) and disregard the
contribution from impurity bound states up to Sec.6. In Sec.3 we determine the
nonmonotony of the relaxation time in the unitary limit. It is shown, that for
the energy dominating regime $p_fv_s\ll \omega $ the relaxation time
$\tau_s(\omega, v_s)$ decreases with increasing superflow in the presence of
the line of nodes of the order parameter, in contrast to the case of superflow
dominating condition $\omega \ll p_fv_s$, when the relaxation time rises.
Analytical calculations are carried out for sufficiently small quantities
$p_fv_s,\ \omega\ll \Delta_{max}$. In Sec.4 the thermal conduction as a
function of the superflow as well as along the applied magnetic field is
analyzed. We examine three important examples of anisotropic gap functions: the
polar state and (1,i) superconducting states for $E_{1g}$ and $E_{2u}$ types of
superconducting pairing in hexagonal crystals (~like $UPt_3$~). We find that in
the temperature dominating region $T_c\sqrt{\frac{\displaystyle
B}{\displaystyle B_{c2}}}\ll T\ll\Delta_{max}$ the thermal conductivity
diminishes with the magnetic field for all three types of pairing considered,
if the magnetic field is applied within the basal plane.  This turns out to be
valid as well for $\kappa_{zz}$ and the magnetic field along $z$-axis (of high
symmetry) for $E_{1g}$ and $E_{2u}$ types of pairing, not for the polar state.
The contribution to the magnetic field dependence of $\kappa$ from the
intervortex space $r\gg\xi$ dominates in this case, which justifies the
semiclassical approximation we use.  For higher fields or lower temperatures
the thermal conductivity always rises with the increasing magnetic field,
although in this case intervortex space dominates only for the magnetic field
oriented within the basal plane. This ensures the nonmonotonous magnetic field
dependence of $\kappa$ (~at least for $E_{2u}$ and $E_{1g}$ models~) under the
condition $\beta T_c\sqrt{\frac{ \displaystyle B_{c1}}{ \displaystyle B_{c2}}}
\le T\ll T_{c}$, where numerical factor $\beta$ may be of the order of unity.
In Sec.5 the low-temperature behavior of the thermal conductivity in zero field
and the magnetic field dependence of $\kappa_{ii}(T,B)$ at low fields  and low
temperatures are studied in clean limit (the $i$-th axis is aligned along the
magnetic field).  We show for the temperature region $\gamma\lesssim T\ll
\Delta_{max}$, that scaling of the thermal conductivity with a single parameter
$x=\frac{T}{T_c}\sqrt{\frac{B_{c2}}{B}}$ at low fields and low temperatures
(both for $i=y$ and for $i=z$) as well as weak low-temperature dependence of
the anisotropy ratio $\kappa_{zz}/\kappa_{yy}$ in zero field, are approximately
valid in clean limit for $(1,i)$-phase of $E_{2u}$-type of pairing.
Qualitatively it is quite close to what it is observed experimentally for
$UPt_3$ in\cite{fl1,fl2}. Under the same conditions $E_{1g}$ model results in
more noticeable deviations from the scaling, and in essential temperature
dependence of the ratio $\kappa_{zz}(T,B=0)/\kappa_{yy}(T,B=0)$. New test is
suggested for discrimination between candidates for the type of pairing in
$UPt_3$, based on the dependence of $\kappa_{zz}$ upon the value of transport
supercurrent flowing in thin films or whyskers along the hexagonal axis in the
absence of the magnetic field. In Sec.6 we describe the behavior of the thermal
conductivity under the condition $p_fv_s, T\lesssim\gamma$, when contribution
from impurity bound states dominates the thermal conductivity. We find in this
limit, that the thermal conductivity monotonously decreases with increasing
condensate flow field (at least for $\gamma\ll \Delta_{max}$) and does not
satisfy the above-mentioned scaling behavior. This is also in agreement with
the experimental results for $UPt_3$.

\section{Basic equations}

Taking into account influence of the magnetic field (satisfying the condition
$B_{c1}\lesssim B\ll B_{c2}$)  on the thermal conductivity, we are interested
in the contribution from large distances from vortex cores ($\xi(T)\lesssim
r$), where the problem can be considered approximately as locally
quasihomogeneous one on the basis of semiclassical approximation\cite{vol3}.
Being justified for those cases, when the contribution from the intervortex
space $\xi\lesssim r$ turns out to be dominating, such an approach simplifies
greatly all analytical considerations permitting to obtain correct results up to
numerical coefficients of the order of unity.  Besides, we assume the
particle-hole symmetry and consider superconducting states to be unitary:
$\hat{\Delta}\hat{\Delta}^\dagger=|\Delta|^2\hat{\sigma}_0$, where
$\hat{\sigma}_0$ is the $2\times2$ unit matrix.  Then the expression for the
thermal conductivity under the applied magnetic field can be written in the
form
$$ \kappa_{ij}= \frac{\displaystyle N_f}{\displaystyle
2}\int_0^{\infty}d\omega\left( \frac{\displaystyle \omega}{\displaystyle
T}\right)^2 \frac{\displaystyle 1}{\displaystyle
\cosh^2\left(\frac{\omega}{2T}\right)} \frac{\displaystyle 1}{\displaystyle
{\rm Im}\ \tilde\omega}
\left <\enspace \frac{\displaystyle
v_{f,i}v_{f,j}}{\displaystyle 2\left( {\rm Re}\ \tilde\omega-
\bbox{v}_f{\bf v}_s\right)}
\times
\qquad \qquad \qquad \qquad \qquad \qquad
\right.
$$
\begin{equation}
\qquad \qquad \qquad
\left.
\times{\rm Re}\left\{
2\sqrt{\left(\tilde\omega-\bbox{v}_f
{\bf v}_s\right)^2-|\tilde\Delta(\bbox{p}_f)|^2}+
\frac{\displaystyle
\left|\tilde\omega-\bbox{v}_f{\bf v}_s\right|^2-
\left(\tilde\omega-\bbox{v}_f{\bf v}_s\right)^2 }{\displaystyle
\sqrt{\left(\tilde\omega-\bbox{v}_f
{\bf v}_s\right)^2-|\tilde\Delta(\bbox{p}_f)|^2}}\right\} \right>_{S_f}
\enspace .
\label{kappaij}
\end{equation}
Here ${\bf v}_s=(1/2)\left({\bbox\nabla}\chi+(2e/c){\bbox A}\right)$ is the
gauge-invariant condensate flow field induced by the magnetic field (see, for
example,\cite{xu}; $\chi$ is the phase of the order parameter) , $\tilde\omega$
and $\tilde\Delta_{\bf{p}}$ are the quasiparticle energy and the order
parameter renormalized by impurities; notation $\left<\dots\right>_{S_f}$
means the averaging over the Fermi surface. Further we ignore contributions to
the quasiparticle group velocity due to the momentum direction dependence of
the order parameter, since for small enough parameter
$\Delta_{max}/\varepsilon_f$ this would lead to a quite small corrections to
the thermal conductivity\cite{mor}. Also we assume below for simplicity a
spherical Fermi surface and introduce the superfluid velocity $v_s={\rm
v}_s/m$, so that under this simplification $\bbox{v}_f{\bf v}_s=
m\bbox{v}_f\bbox{v}_s=\bbox{p}_f\bbox{v}_s$. The branch of square root function
in (\ref{kappaij}) should be chosen to have nonnegative imaginary part.

Considering sufficiently clean superconductors in the temperature region
$\gamma\lesssim T$, one can disregard the contribution from impurity bound
states. Then Eq.(\ref{kappaij}) reduces to
\begin{equation}
\frac{\displaystyle\kappa_{ij}}{\displaystyle\kappa_N(T_c)}=
\frac{\displaystyle 9}{\displaystyle 2\pi^2T_c}\int_0^{\infty}d\omega\left(
\frac{\displaystyle \omega}{\displaystyle T}\right)^2
\frac{\displaystyle 1}{\displaystyle \cosh^2\left(\frac{\omega}{2T}\right)}
\frac{\tau_s(\omega, v_s)}{\tau_N}I_{ij}(\omega, v_s) \enspace .
\label{thc}
\end{equation}
Here ${\kappa_N(T_c)}$ is the thermal conductivity in the normal state at
$T=T_c$; $\tau_s(\omega)$ and $\tau_N$ are relaxation times for quasiparticle
scattering on impurities in superconducting
and normal-metal states respectively. The quantity $I_{ij}(\omega, v_s)$ in the
case of spherical Fermi surface is defined as follows
\begin{equation}
I_{ij}(\omega, v_s)=\int\limits_{|\Delta({\bf p}_f)|\le
|\omega-{\bf p}_f{\bf v}_s|}
\frac{\displaystyle d\Omega}{\displaystyle 4\pi}
\frac{\displaystyle p_ip_j}{\displaystyle p^2}
\sqrt{1-\frac{\displaystyle|\Delta(\bbox{p}_f)|^2}{
\displaystyle\left(\omega-\bbox{p}_f\bbox{v}_s\right)^2}} \enspace .
\label{I}
\end{equation}

Assuming particle-hole symmetry, one can represent the relaxation time in
the Born approximation in the form:
\begin{equation}
\frac{\displaystyle \tau_s(\omega)}{\displaystyle \tau_N}= \frac{
\displaystyle 1}{\displaystyle {\rm Im} G_0(\omega)} \enspace ,
\label{tauborn}
\end{equation}
while in the unitarity limit the corresponding expression is
\begin{equation}
\frac{\displaystyle \tau_s(\omega)}{\displaystyle \tau_N}=
\frac{\displaystyle 1}{\displaystyle{\rm Im}
\frac{\displaystyle G_0(\omega)}{\displaystyle G_1^2(\omega)-G_0^2(\omega)}}
\enspace .
\label{tauuni}
\end{equation}

The quantities $G_0(\omega)$, $G_1(\omega)$  are
\begin{equation}
G_0(\omega)=i\int\frac{\displaystyle d\Omega}{\displaystyle 4\pi}
\frac{\displaystyle \omega-\bbox{p}_f\bbox{v}_s}{\displaystyle
\sqrt{\left(\omega-\bbox{p}_f\bbox{v}_s\right)^2-|\Delta(\bbox{p}_f)|^2}}
\enspace ,
\label{G0}
\end{equation}
\begin{equation}
G_1(\omega)=i\int\frac{\displaystyle d\Omega}{\displaystyle 4\pi}
\frac{\displaystyle \Delta(\bbox{p}_f)}{\displaystyle
\sqrt{\left(\omega-\bbox{p}_f\bbox{v}_s\right)^2-|\Delta(\bbox{p}_f)|^2}}
\enspace .
\label{G1}
\end{equation}
In the absence of the superfluid velocity above expressions are reduced to
well-known relations\cite{hir1}. For the chosen branch of square root function
the real part of the integrand in (\ref{G0}) (that is the density of states) is
a nonnegative quantity.

\section{Relaxation time in the unitarity limit}

\subsection{Magnetic field, aligned along $z$-axis}

Let us consider superconducting phases with order parameters having line  of
nodes in the equatorial plane on a (spherical) Fermi surface, and the applied
magnetic field satisfying the condition $B_{c1}< B\ll B_{c2}$ and aligned along
the crystalline axis of high symmetry ($z$-axis).  As particular examples,
there may be the polar phase and $(1,i)$-superconducting phases of $E_{1g}$-
and $E_{2u}$-types of pairing in hexagonal superconductors.  Last two phases
are widely discussed as candidates for the type of superconducting pairing
taking place in the heavy fermion superconductor
$UPt_3$\cite{sigueda,sauls,joynt}. All these particular order parameters change
their signs under the reflection across the equatorial plane on the Fermi
surface:  $\Delta(\pi-\theta)=-\Delta(\theta)$. Due to this property  function
$G_1(\omega)$ is equal to zero for the magnetic field parallel to $z$-axis.

The applied magnetic field influences the relaxation time mainly due to its
dependence upon the superflow, induced by the field and aligned in the
particular case parallel to the basal plane.  In calculating the dependence of
the relaxation time $\tau_s(\omega)$ upon the superflow under the condition
$p_fv_s,\,\omega\ll\Delta_{max}$, it is essential that only narrow regions near
nodes of the order parameter on the Fermi surface govern the dependence of
$G_0(\omega, v_s)-{\rm Re}G_0(\omega, v_s=0)$ upon the superfluid velocity. At
the same time the quantity ${\rm Re}G_0(\omega, v_s=0)$, generally speaking, is
formed by the whole Fermi surface.  We assume that the order parameter near the
line of nodes ($|\frac{\pi}{2}-\theta|\ll 1$) may be represented as
$|\Delta|=\Delta_1|\frac{\pi}{2}-\theta|$, while near the point node on one of
the poles ($|\theta|\ll 1$) it has the form $|\Delta|=\Delta_2|\theta|^n$,\,
$n=1,2$. Calculations show, that if one keeps terms up to $p_fv_s/\Delta_{1,2}$
in the expansion in powers of the superfluid velocity, taking into account as
well the first power of small parameter $\omega/\Delta_{1,2}$, then
contributions from the line of nodes and quadratic points are essential. At the
same time contributions from linear points may be neglected as compared to ones
from the line of nodes.  As a result we obtain the following expressions for
the relaxation time in the unitarity limit
$$ \frac{\displaystyle
\tau_s(\omega)}{\displaystyle \tau_N}=
\frac{\displaystyle\pi\omega}{\displaystyle 2\Delta_1}\left(1+
\frac{\displaystyle \Delta_1}{\displaystyle 2\Delta_2}\right) +
\frac{\displaystyle 2\omega}{\displaystyle \pi\Delta_1\left(1+
\frac{\displaystyle \Delta_1}{\displaystyle 2\Delta_2}\right)}
\left[\frac{\displaystyle \Delta_1}{\displaystyle \omega} {\rm
Re}G_0(\omega,v_s=0)+ \qquad\qquad\qquad\qquad\qquad \right.
$$
\begin{equation}
\left.
+\ln\left(\frac{\displaystyle 2\omega}{\displaystyle
e\left(\omega+\sqrt{\omega^2-(p_fv_s)^2}\right)}\right)+
\sqrt{1-\left(\frac{\displaystyle p_fv_s}{\displaystyle\omega}\right)^2} \
\right]^2 \enspace , \qquad
p_fv_s\le \omega \ll \Delta_{max} \enspace ,
\label{tau1}
\end{equation}

$$
\frac{\displaystyle \tau_s(\omega)}{\displaystyle \tau_N}=
\frac{\displaystyle p_fv_s }{\displaystyle \Delta_1}\left(
\sqrt{1-\left(\frac{\displaystyle \omega}{\displaystyle p_fv_s }\right)^2}+
\frac{\displaystyle \omega}{\displaystyle p_fv_s }
\sin^{-1}\left(\frac{\displaystyle \omega}{\displaystyle p_fv_s }\right)\right)
+\frac{\displaystyle \pi \omega }{\displaystyle 4\Delta_2 }+
\qquad\qquad\qquad\qquad\qquad
$$
\begin{equation}
+\frac{\displaystyle \omega^2
\left(\frac{\displaystyle \Delta_1}{\displaystyle \omega} {\rm Re}
G_0(\omega,v_s=0) +\ln\left(\frac{\displaystyle 2\omega}{\displaystyle
ep_fv_s }\right)\right)^2}{\displaystyle p_fv_s \Delta_1\left(
\sqrt{1-\left(\frac{\displaystyle \omega}{\displaystyle p_fv_s }\right)^2}+
\frac{\displaystyle \omega}{\displaystyle p_fv_s }
\sin^{-1}\left(\frac{\displaystyle \omega}{\displaystyle p_fv_s }\right)
+\frac{\displaystyle \pi \omega \Delta_1}{\displaystyle 4p_fv_s \Delta_2 }
\right)}
\enspace , \qquad  \omega\le p_fv_s \ll \Delta_{max} \enspace .
\label{tau2}
\end{equation}

The most important feature of the dependence upon the superflow, describing by
Eqs.(\ref{tau1}), (\ref{tau2}), is its nonmonotonous behavior. Indeed, one can
easily see under the conditions $p_fv_s\ll \omega \ll \Delta_{max}$, that the
relaxation time decreases with the increasing superfluid velocity according to
the relation
$$
\frac{\displaystyle \tau_s(\omega,v_s)}{\displaystyle \tau_N}=
\frac{\displaystyle\pi\omega}{\displaystyle 2\Delta_1}\left(1+
\frac{\displaystyle \Delta_1}{\displaystyle 2\Delta_2}\right) +
\frac{\displaystyle 2}{\displaystyle \pi\left(1+
\frac{\displaystyle \Delta_1}{\displaystyle 2\Delta_2}\right)}
\left(\frac{\displaystyle \Delta_1}{\displaystyle \omega}\left({\rm Re}
G_0(\omega,v_s=0)\right)^2- \quad\qquad\qquad\qquad\qquad
\right.
$$
\begin{equation}
\left.
\qquad\qquad\qquad\qquad\qquad\qquad\qquad\qquad\qquad\qquad\qquad\qquad
-\frac{\displaystyle (p_fv_s)^2}{\displaystyle 2\omega^2}{\rm Re}
G_0(\omega,v_s=0)\right) \enspace .
\label{tau3}
\end{equation}

At the same time under the opposite condition $\omega\ll p_fv_s\ll
\Delta_{max}$ we see from (\ref{tau2}), that $\tau_s/\tau_N=p_fv_s/\Delta_1$
increases linearly with the parameter $p_fv_s$.

Expressions (\ref{tau1})-(\ref{tau3}) describe directly the relaxation time for
$(1,i)$ superconducting state of $E_{2u}$-type of pairing, for which the order
parameter has the line of nodes on the equator and quadratic point nodes on the
poles. Results for $E_{1g}$-type of pairing (~line of nodes on the equator and
linear point nodes on the poles~) and for the polar phase (~nodes are only
on the line on the equator~) follow from (\ref{tau1})-(\ref{tau3}) in the limit
$\Delta_2\rightarrow \infty$.

In calculating ${\rm Re}G_0(\omega,v_s=0)$ one should describe the particular
form of the order parameter all over the Fermi surface. In the particular case
of polar phase $\Delta=\Delta_1\cos\theta$ explicit integration in (\ref{G0})
results in the following expression
\begin{equation}
{\rm Re}G_0(\omega,v_s=0)=\frac{\displaystyle \omega}{\displaystyle \Delta_1}
\ln\left( \frac{\displaystyle \Delta_1}{\displaystyle \omega}+
\sqrt{\left(\frac{\displaystyle \Delta_1}{\displaystyle \omega}\right)^2
-1}\right) \enspace .
\label{reg0}
\end{equation}

Substituting (\ref{reg0}) into  (\ref{tau1}) and (\ref{tau2}), one gets that at
the point of minimum of the relaxation time the superfluid velocity satisfies
the condition $p_fv_s\sim \omega\ln\left(\frac{\displaystyle\Delta_1
}{\displaystyle \omega}\right)$.  Under the condition $p_fv_s\ll \omega \ll
\Delta_{1}$ relaxation time decreases with increasing parameter $p_fv_s$:
\begin{equation}
\frac{\displaystyle \tau_s(\omega,v_s)}{\displaystyle \tau_N}=
\frac{\displaystyle 2\omega}{\displaystyle \pi\Delta_1}\left[
\frac{\displaystyle \pi^2}{\displaystyle 4} + \ln^2\left(\frac{\displaystyle
2\Delta_1}{\displaystyle \omega}\right) -\frac{\displaystyle
(p_fv_s)^2}{\displaystyle 2\omega^2} \ln\left(\frac{\displaystyle
2\Delta_1}{\displaystyle \omega}\right) \right] \enspace ,
\label{pol1tau}
\end{equation}
while in the opposite case $\omega \ll p_fv_s\ll \Delta_{1}$ we have
\begin{equation}
\frac{\displaystyle \tau_s(\omega,v_s)}{\displaystyle \tau_N}=
\frac{\displaystyle p_fv_s}{\displaystyle \Delta_1}\left[
1+\left(\frac{\displaystyle \omega}{\displaystyle p_fv_s}\right)^2
\ln^2\left(\frac{\displaystyle 4\Delta_1}{\displaystyle ep_fv_s}\right) \right]
\enspace .
\label{pol2tau}
\end{equation}
If $\omega\ln\left(\frac{\displaystyle\Delta_1 }{\displaystyle\omega}\right)
\lesssim p_fv_s$, expression (\ref{pol2tau}) increases with increasing
$p_fv_s$.

One can see, that nonmonotonous behavior of the relaxation time with superfluid
velocity takes place in the unitary limit due to
different behaviors of ${\rm Re}G_0$ and ${\rm Im}G_0$. At sufficiently low
energy the density of states (and, hence, ${\rm Im}G_0$) increases with
increasing magnetic field, while ${\rm Re}G_0$ turns out to decrease.
In the case in question the relaxation time is described by the
relation\cite{bah1}
\begin{equation}
\frac{\displaystyle \tau_s(\omega)}{\displaystyle \tau_N}= {\rm Im}G_0(\omega)
+\frac{\displaystyle\left({\rm Re}G_0(\omega)\right)^2}{\displaystyle{\rm Im}
G_0(\omega)}
\label{tauun1}
\end{equation}
and may manifest nonmonotonous magnetic field dependence (see Fig.\ 1).

\subsection{Magnetic field parallel to the basal plane}

Nonmonotony of the relaxation time as a function of the applied magnetic field
may take place for various orientations of the field, due to the same reasons
as for the field parallel to the $z$-axis. In this subsection we present
shortly some results on magnetic field dependence of relaxation time for the
field lying within the basal plane for polar phase
$\left(\Delta(\bbox{p}_f)=\Delta_p(\theta)\right)$, $(1,i)$-phase of
$E_{1g}$-representation of $D_{6h}$ point group
$\left(\Delta(\bbox{p}_f)=\Delta_{1g}(\theta)\exp(i\varphi)\right)$, and
$(1,i)$-phase of $E_{2u}$-representation
$\left(\Delta(\bbox{p}_f)=\Delta_{2u}(\theta)\exp(2i\varphi)\right)$.
Spherical angles $\theta$ and $\varphi$ characterize here momentum directions
on a (spherical) Fermi surface, and functions $\Delta_p(\theta)$,
$\Delta_{1g}(\theta)$, $\Delta_{2u}(\theta)$ are supposed to be real.

Let $y$-axis be parallel to the magnetic field. We consider some fixed spatial
point far enough from the vortex core, where induced superfluid velocity, lying
in $xz$-plane, constitutes angle $\theta_0$ with $z$-axis.  Although for the
field orientation within the basal plane $G_1(\omega)$ differs from zero, the
analysis shows that this quantity may be neglected as compared to
$G_0(\omega)$, at least for the types of pairing considered and under the
condition $\omega,\,p_fv_s\ll\Delta_{max}$. This allows one to use
Eq.(\ref{tauun1}) in calculating $\tau_s(\omega)$.  Assuming  the same
relations for the order parameter near the line of nodes
($|\Delta|=\Delta_1|\frac{\pi}{2}-\theta|$) and point node
($|\Delta|=\Delta_2|\theta|^n$,\ $n=1,2$) as earlier, we get again (within the
same accuracy)  that the relaxation time diminishes with increasing superfluid
velocity under the conditions $p_fv_s\ll\omega\ll\Delta_{max}$:
$$
\frac{\displaystyle \tau_s(\omega,v_s)}{\displaystyle \tau_N}=
\frac{\displaystyle\pi\omega}{\displaystyle 2\Delta_1}\left(1+
\frac{\displaystyle \Delta_1}{\displaystyle 2\Delta_2}\right) +
\frac{\displaystyle 2}{\displaystyle \pi\left(1+
\frac{\displaystyle \Delta_1}{\displaystyle 2\Delta_2}\right)}
\left[\frac{\displaystyle \Delta_1}{\displaystyle \omega}\left({\rm Re}
G_0(\omega,v_s=0)\right)^2- \quad\qquad\qquad\qquad\qquad
\right.
$$
\begin{equation}
\left.
\qquad\qquad\qquad\qquad\qquad\qquad\qquad\qquad\quad
-\left(1+\frac{\displaystyle\Delta_1}{\displaystyle \Delta_2}\right)
\frac{\displaystyle (p_fv_s)^2\cos^2\theta_0}{\displaystyle 2\omega^2}
{\rm Re}G_0(\omega,v_s=0)\right] \enspace .
\label{tau3p}
\end{equation}

In the opposite case $\omega\ll p_fv_s\ll\Delta_{max}$ relaxation time
increases with $v_s$:
\begin{equation}
\frac{\displaystyle \tau_s(v_s)}{\displaystyle \tau_N}=
p_fv_s\left(\frac{\displaystyle|\sin\theta_0|}{\displaystyle\Delta_1} +
\frac{\displaystyle\pi|\cos\theta_0|}{\displaystyle 4\Delta_2}\right)\enspace .
\label{tau4p}
\end{equation}

Eqs.(\ref{tau3p}), (\ref{tau4p}) describe the relaxation time for $(1,i)$
superconducting state of $E_{2u}$-type of pairing.  As it was in the case of
Eqs.(\ref{tau1})-(\ref{tau3}), results for $E_{1g}$-type of pairing and for the
polar phase follow from (\ref{tau3p}), (\ref{tau4p}) in the limit
$\Delta_2\rightarrow \infty$.  For the polar phase we obtain under the
condition $\omega\ll\Delta_1$ the explicit expression ${\rm
Re}G_0(\omega,v_s=0)= (\omega/\Delta_1)\ln(2\Delta_1/\omega)$. Then
Eq.(\ref{tau3p}) reduces to
\begin{equation}
\frac{\displaystyle \tau_s(\omega,v_s)}{\displaystyle \tau_N}=
\frac{\displaystyle 2\omega}{\displaystyle \pi\Delta_1}\left[
\frac{\displaystyle \pi^2}{\displaystyle 4}+\ln^2\left(\frac{\displaystyle
2\Delta_1}{\displaystyle \omega}\right)-
\frac{\displaystyle1}{\displaystyle2}\left(\frac{\displaystyle
p_fv_s\cos\theta_0}{\displaystyle \omega}\right)^2\ln\left(\frac{\displaystyle
2\Delta_1}{\displaystyle \omega}\right)\right] \enspace .
\label{rtpp}
\end{equation}

So, magnetic field dependence of the relaxation time for the magnetic field
oriented within the basal plane, turns out to be qualitatively quite close to
the one for magnetic field orientation along the $z$-axis.

\section{Thermal Conductivity}

In this section we consider the electronic thermal conduction along the
direction of the applied magnetic field. Let the magnetic field be aligned
firstly along the $z$-axis, so that the supercurrent flows parallel to the
basal plane.

For low temperatures and small superfluid velocities satisfying the conditions
$T,\ p_fv_s\ll\Delta_{max}$\ , there are two characteristic regions $p_fv_s\ll
T\ll\Delta_{max}$ and $T\ll p_fv_s\ll\Delta_{max}$, where dependences of the
thermal conductivity upon $v_s$ may differ essentially.  While within the
region $T\ll p_fv_s\ll\Delta_{max}$ the thermal conductivity always rises with
increasing $v_s$, it may be possible both increase or decrease under the
condition $p_fv_s\ll T\ll\Delta_{max}$.  Thus, nonmonotonous behavior of the
relaxation time may result in respective nonmonotonous dependence of the
thermal conductivity upon the superflow (and eventually upon the magnetic
field).  However, nonmonotony is not a inevitable consequence for the thermal
conductivity, which would be valid for any types of pairing in the presence of
such a behavior of the relaxation time.  The point is that the quantity
$I_{ij}(\omega, v_s)$, which forms the behavior of $\kappa$ along with the
relaxation time  (see (\ref{thc})), usually turns out to be monotonously
increasing function of $v_s$.  As a result, the product $\tau_s(\omega,
v_s)I_{ij}(\omega, v_s)$, containing nonmonotonous function $\tau_s(\omega,
v_s)$, may manifest both monotonous or nonmonotonous behavior, depending on the
particular behaviors of $\tau_s(\omega, v_s)$ and $I_{ij}(\omega, v_s)$.

For sufficiently low temperatures $T\ll p_fv_s\ll\Delta_{max}$ (or, in other
words, for large enough superfluid flow), it follows from (\ref{thc}) in the
first approximation
\begin{equation}
\frac{\displaystyle\kappa_{ij}}{\displaystyle\kappa_N(T_c)}=
\frac{\displaystyle 3T}{\displaystyle T_c}
\frac{\tau_s(\omega=0, v_s)}{\tau_N}I_{ij}(\omega=0, v_s) \enspace .
\label{thclt}
\end{equation}
The possibility to disregard the contribution from impurity induced bound
states in this case is justified under the condition $p_fv_s\gg \gamma$.

One can show, that the contribution to $I_{zz}(\omega=0, v_s)$ from narrow
region around the line of nodes is $(p_fv_s)^3/12\Delta_1^3$. Linear point
nodes do not contribute to $I_{zz}(\omega=0, v_s)$, so that Eq.(\ref{thclt}) is
not the appropriate approximation for the case of the order parameter having
only linear point nodes. It is important, however, that the contribution from
linear point nodes in this limit is evidently less than from the line of nodes.
At the same time the contribution from the quadratic point nodes dominates
being equal to $(p_fv_s)^2/6\Delta_2^2$.  Furthermore, since ${\rm
Re}G_0(\omega=0, v_s=0) =0$, we find from (\ref{tau2}):\ $\tau_s(\omega=0,
v_s)/\tau_N=p_fv_s/\Delta_1$. The contribution to $\tau_s(\omega=0,
v_s)/\tau_N$ from quadratic point node is $(p_fv_s)^2/2\Delta_2^2$, which can
be disregarded as compared to the above value of $\tau_s(\omega=0, v_s)/\tau_N$
formed by the line of nodes.  Subsequently, under the condition $T\ll
p_fv_s\ll\Delta_{max}$ we find in the case of polar phase and for (1,i)-phase
of $E_{1g}$-type of paring:
\begin{equation}
\frac{\displaystyle\kappa_{zz}}{\displaystyle\kappa_N(T_c)}=
\frac{\displaystyle T}{\displaystyle 4T_c}\left(\frac{\displaystyle
p_fv_s}{\displaystyle \Delta_1}\right)^4 \ .
\end{equation}
For (1,i)-phase of $E_{2u}$-type of paring it follows, that
\begin{equation}
\frac{\displaystyle\kappa_{zz}}{\displaystyle\kappa_N(T_c)}=
\frac{\displaystyle T}{\displaystyle 2T_c}\frac{\displaystyle
(p_fv_s)^3}{\displaystyle \Delta_1\Delta_2^2} \ .
\end{equation}

These higher power law dependences upon the local superfluid velocity result
after averaging over the intervortex space in linear dependence upon the
magnetic field, formed by the lower border $r\sim\xi$ of the intervortex space
considered. Indeed, superfluid velocity ${v}_s({r})$ depends upon the
distance from the vortex core and in the simplest case of circular
supercurrents one has ${v}_s(r)=1/2m_er$ for $\xi\ll r\ll \lambda$.  Then we
obtain after spatial averaging:
\begin{equation}
\kappa_{zz}(T,H)\simeq\frac{\displaystyle 1}{\displaystyle 4}
\left(\frac{\displaystyle p_fv_s(\xi)}{\displaystyle \Delta_1}\right)^4
\frac{\displaystyle BT}{\displaystyle B_{c2}T_c}\kappa_N(T_c)
\label{thcp1}
\end{equation}
for polar phase and for (1,i)-phase of $E_{1g}$-type of paring,
and
\begin{equation}
\kappa_{zz}(T,H)\simeq
\frac{\displaystyle (p_fv_s(\xi))^3}{\displaystyle \Delta_1\Delta_2^2}
\frac{\displaystyle BT}{\displaystyle B_{c2}T_c}\kappa_N(T_c)
\label{thcp2}
\end{equation}
for (1,i)-phase of $E_{2u}$-type of paring.

Since regions close to vortex cores ($r\sim\xi$) dominate in spatial averaging
of the thermal conductivity under the particular conditions $T\ll p_fv_s$,
${\bf B}\| Oz$, the approximation of locally quasiuniform superflow, generally
speaking, is not directly applied to this case. In addition, the relation
$p_fv_s(\xi)\sim T_c$ guarantees validity of inequality $T\ll p_fv_s$ for all
$T\ll T_c$ and, from the other side, violates the condition $p_fv_s\ll
\Delta_{max}$, which leads to the conclusion about rough qualitative character
of the particular results derived above. Strictly speaking, in this particular
case it is necessary to check the contribution to $\kappa$ not only from the
delocalized quasiparticles, but from quasiparticles bound to vortex cores as
well. That problem is, however, beyond the present article. It is important,
that in any case the thermal conductivity increases with the increasing
magnetic field in the limit considered.  This is, fortunately, the only
particular example in the article, where the intervortex space doesn't
dominate.  Furthermore, it is known for $s$-wave superconductors, that
contribution from bound states (~localized within vortex cores~) to the
low-temperature behavior of the thermal conductivity along the magnetic field,
is negligibly small as compared to the normal state thermal conductivity
$\kappa_N$\cite{vinen}.  In the presence of nodes of the order parameter
extended (scattering) low-energy quasiparticle states contribute much greater.
Indeed, as $p_fv_s(\xi)/\Delta_{max}\sim 1$, we get from Eqs.(\ref{thcp1}),
(\ref{thcp2}) $\kappa_{zz}(T)\sim (B/B_{c2})\kappa_N(T) $.  This estimate is
quite close to that observed experimentally for $UPt_3$ for low
temperatures\cite{fl1,fl2} and about on two orders greater, than the one
realized for conventional superconductors\cite{vinen}. Also we note, that for
the field within the basal plane large distances from the vortex cores dominate
for any relation between $T$ and $p_fv_s$  (see below) ensuring the
applicability of the semiclassical approximation in that case both in
temperature dominating regime $T\gg p_fv_s$ and under the superflow dominating
condition $T\ll p_fv_s$ for three types of pairing discussed. At last, for Born
scatterers the situation changes and in spatial averaging in the superflow
dominating limit large distances from vortex cores dominate both for
$\kappa_{zz}$ and the field along the $z$-axis and for $\kappa_{yy}$ and the
field along the $y$-axis.

Let us discuss now the other case $p_fv_s\ll T\ll\Delta_{max}$.  It is
worth noting that this condition can't be valid close to vortex cores (for
$r\sim\xi$). However, as we show below, under the condition $p_fv_s\ll T\ll
\Delta_{max}$ fairly large distances from vortex cores (of the order of the
intervortex distance $R\sim\xi\sqrt{\frac{\displaystyle B_{c2}}{\displaystyle
B}}\gg\xi$) may dominate in the spatially averaged thermal conductivity.

Considering firstly the thermal conductivity for the polar phase and
substituting $\Delta=\Delta_1\cos\theta$ into (\ref{I}), we obtain under the
condition $\omega,\ p_fv_s\ll\Delta_1$\ :
\begin{equation}
I_{zz}(\omega, v_s)=\frac{\displaystyle 1}{\displaystyle 16\Delta_1^3}
\left\{\begin{array}{lcl}
\omega\left[2\omega^2+3(p_fv_s)^2\right]
\sin^{-1}\left(\frac{\displaystyle \omega}{\displaystyle p_fv_s }\right)
+\left[3\omega^2+
\right.
\\
\left.
\quad
+2(p_fv_s)^2\right]\sqrt{(p_fv_s)^2-\omega^2}
-\frac{\displaystyle 2}{\displaystyle 3 }\left[(p_fv_s)^2-
\omega^2\right]^{3/2} & , & \omega\le p_fv_s \enspace ,\\
&&\\
\pi\omega^3+ \frac{\displaystyle 3\pi}{\displaystyle 2}\omega(p_fv_s)^2
& , &   \omega\ge p_fv_s \enspace . \\
\end{array}
\right.
\label{Ip}
\end{equation}

For $\omega\lesssim p_fv_s$ it follows from (\ref{Ip}), (\ref{tau2}), that
$I(\omega, v_s)\lesssim\left(\frac{\displaystyle p_fv_s}{\displaystyle
\Delta_1}\right)^3$, \ $\tau_s\sim \left(\frac{\displaystyle
p_fv_s}{\displaystyle\Delta_1}\right)$. Taking account of these estimates, we
find, that under the condition $p_fv_s\ll T\ll \Delta_1$  the contribution to
the thermal conductivity (\ref{thc}) from the integration over the region
$\omega\lesssim p_fv_s$ turns out to be $\lesssim\frac{\displaystyle
\left(p_fv_s\right)^7}{\displaystyle \Delta_1^4 T_cT^2}$\ .  Since we are
interested in contributions to $\kappa$ up to $(p_fv_s)^2$, one should consider
the integration over $\omega\ge p_fv_s$ and substitute into (\ref{thc})
equations (\ref{pol1tau}), (\ref{Ip}):
$$
\frac{\displaystyle\kappa_{zz}}{\displaystyle\kappa_N(T_c)}=
\frac{\displaystyle 161.74\, T^5}{\displaystyle
T_c\Delta_1^4}
\left[\ln^2\left(\frac{\displaystyle 2\Delta_1}{\displaystyle
T}\right)-
1.63\ln\left(\frac{\displaystyle 2\Delta_1}{\displaystyle T}\right)+
3.16\right]+
\qquad \qquad \qquad \qquad \qquad\qquad\quad
$$
\begin{equation}
\qquad \qquad \qquad \qquad\qquad\qquad\quad
+\frac{\displaystyle 7.77\, T^3 (p_fv_s)^2}{\displaystyle T_c\Delta_1^4}
\left[\ln^2\left(\frac{\displaystyle
2\Delta_1}{\displaystyle T}\right)-1.67\ln\left(\frac{\displaystyle
2\Delta_1}{\displaystyle T}\right)+3.18\right] \enspace .
\label{thcpol2}
\end{equation}
We see from (\ref{thcp1}) and (\ref{thcpol2}), that $\kappa_{zz}$ for polar
state is a monotonously increasing function of the superfluid velocity.
According to (\ref{thcpol2}) the thermal conductivity for polar state in the
presence of magnetic field roughly $\propto T^3$, while in the absence of
magnetic field $\propto T^5$.

By contrast, the thermal conductivity of $(1,i)$-phases for $E_{1g}$ and
$E_{2u}$ types of pairing turns out to be nonmonotonous one.  The point is
that the contribution to $I_{zz}$ from linear point nodes is approximately
$\omega^2/3\Delta_2^2$, while from quadratic point nodes is
$\pi\omega/8\Delta_2$. These terms dominate as compared to the contribution
from line of nodes. This means that magnetic field dependence of the thermal
conductivity is formed in the cases in question entirely by the relaxation
time:
\begin{equation}
\left.
\frac{\displaystyle\kappa_{zz}}{\displaystyle\kappa_N(T_c)}=
\frac{\displaystyle\kappa_{zz}}{\displaystyle\kappa_N(T_c)}\right\vert_{v_s=0}-
\frac{\displaystyle 9(p_fv_s)^2}{\displaystyle
\pi^3T_c\left(2\Delta_2+\Delta_1\right)}\int_0^{\infty}
\frac{\displaystyle \omega d\omega}{\displaystyle
\cosh^2\left(\frac{\omega}{2}\right)}{\rm Re}G_0(T\omega,0)\left\{
\begin{array}{lll}
\frac{\displaystyle T\omega}{\displaystyle 3\Delta_2}& , \ &
for E_{1g}\\
\frac{\displaystyle \pi}{\displaystyle 8}& , \ &for E_{2u}\\
\end{array}
\enspace .
\right.
\label{thce}
\end{equation}

This nonmonotony is seen in Fig.\ 2  where we plot the thermal conductivity for
three particular types of pairing and various values of temperature as a
function of the superfluid velocity\cite{f}.

After averaging over the intervortex space, where expressions (\ref{thcpol2}),
(\ref{thce}) are valid, we obtain
\begin{equation}
\left.
\frac{\displaystyle\kappa_{zz}}{\displaystyle\kappa_N(T_c)}=
\frac{\displaystyle\kappa_{zz}}{\displaystyle\kappa_N(T_c)}\right\vert_{B=0}\pm
A_{zz}(T)\frac{\displaystyle B}{\displaystyle B_{c2}}
\ln\left(\frac{\displaystyle B_{c2}}{\displaystyle B}\right).
\label{thca}
\end{equation}
Sign plus corresponds here to the case of polar phase, while sign minus - to
$E_{1g}$ and $E_{2u}$ types of pairing.  Since ${\rm Re}G_0(T\omega,0)$ is
roughly proportional to $T$ (neglecting the logarithmic factors; see below
Eq.(\ref{g0})), positive dimensionless function $A_{zz}(T)$ turns out to have a
linear temperature dependence for $E_{2u}$ model, a quadratic temperature
dependence for $E_{1g}$, while for the polar phase $A_{zz}(T)$ is roughly a
cubic function of temperature (see Eq.(\ref{thcpol2}) as well). Large
logarithmic factor in magnetic field dependence (\ref{thca}) is formed by
contributions from sufficiently large distances from vortex cores (of the order
of intervortex distance $R$). This justifies the approach we use, which is
based on the consideration of quasihomogeneous superconducting intervortex
space and valid for large distances from vortex cores $r\gg \xi$.

So, in spatial averaging of the thermal conductivity the intervortex space,
generally speaking, is divided into two parts. The first one is defined by the
condition $T\lesssim p_fv_s(r)$ and contributes to the monotonously increasing
thermal conductivity with the magnetic field. The other part of the intervortex
space $p_fv_s(r)\lesssim T$ includes fairly large distances from vortex cores.
For $E_{1g}$ and $E_{2u}$ representations it contributes to decreasing thermal
conductivity with increasing magnetic field and turns out to be essential (that
is may compete with the contribution from the first part) only under the
condition $p_fv_s(R)\ll T$.  This condition reduces to the inequality $T\gg
T_c\sqrt{\frac{ \displaystyle B}{ \displaystyle B_{c2}}}$ which has to be
satisfied ensuring the possibility for the nonmonotonous magnetic field
dependence of the thermal conductivity.  Thus, in the presence of nonmonotony
the minimum of $\kappa(B)$ has to be shifted gradually to lower fields with
decreasing temperature. Below some characteristic temperature $\sim\beta
T_c\sqrt{\frac{ \displaystyle B_{c1}}{ \displaystyle B_{c2}}}$\ this
nonmonotony disappears at all. Rough estimates show that numerical factor
$\beta$ may be of the order of unity. This is in agreement with that observed
experimentally in $UPt_3$\cite{fl1,fl2}. Naturally, the relation
$\beta T_c\sqrt{\frac{ \displaystyle B_{c1}}{ \displaystyle B_{c2}}}\lesssim
T\ll T_c$ can be satisfied only for large enough value of the
Ginzburg-Landau parameter as this takes place, in particular, in $UPt_3$.

Let now the applied magnetic field lie in the basal plane along $y$-axis.
While the relaxation time behavior is quite similar for both orientations of
the magnetic field, it is not always the case for $I_{ij}$. As a consequence, a
qualitative behavior of the thermal conduction along the magnetic field for
$B\| Oz$ and $B\| Oy$ may differ as well. In particular, we show below for the
field orientation $B\parallel Oy$, that nonmonotonous magnetic field dependence
of the thermal conductivity $\kappa_{yy}$ takes place for all three types of
pairing in question, including the polar phase (in contrast to $\kappa_{zz}$
for the orientation ${\bf B}\| Oz$).

Under conditions $\omega,\, p_fv_s\ll \Delta_{max}$ we get, that the main
contribution to $I_{yy}$ comes only from the narrow region of the line of nodes
for all types of pairing considered (including $E_{2u}$~--~representation
with quadratic point nodes on the poles of the Fermi surface):
\begin{equation}
I_{yy}=\left\{\begin{array}{llll}
\frac{\displaystyle \pi\omega}{\displaystyle 8\Delta_1}&,\hspace{0.5cm}
p_fv_s|\sin\theta_0|\le\omega &&\\
\multicolumn{2}{l}
{\frac{\displaystyle p_fv_s|\sin\theta_0|}{\displaystyle 4\Delta_1}
\left[\frac{\displaystyle 2}{\displaystyle 3}\left(1-\frac{\displaystyle
\omega^2}{\displaystyle p^2_fv^2_s\sin^2\theta_0}\right)^{3/2}+
\frac{\displaystyle \omega}{\displaystyle p_fv_s|\sin\theta_0|}\sin^{-1}
\frac{\displaystyle \omega}{\displaystyle p_fv_s|\sin\theta_0|}+\right.}
\\
\left.
\quad\qquad\qquad\qquad
+\frac{\displaystyle \omega^2}{\displaystyle p^2_fv^2_s\sin^2\theta_0}
\sqrt{1-\frac{\displaystyle \omega^2}{\displaystyle p^2_fv^2_s\sin^2\theta_0}}
 \quad \right] & , \hspace{0.5cm} p_fv_s|\sin\theta_0|\ge \omega  & &
\enspace .\\
\end{array}
\right.
\label{iyy}
\end{equation}

As it follows from (\ref{iyy}),
$I_{yy}(\omega=0,v_s)=p_fv_s|\sin\theta_0|/6\Delta_1$. Substituting this value
together with (\ref{tau4p}) into (\ref{thclt}), we obtain the thermal
conductivity at low temperatures $T\ll p_fv_s\ll\Delta_{max}$ for
$(1,i)$-phase of $E_{2u}$-type of pairing:
\begin{equation}
\frac{\displaystyle\kappa_{yy}}{\displaystyle\kappa_N(T_c)}=(p_fv_s)^2
\frac{\displaystyle T}{\displaystyle2T_c}\left(
\frac{\displaystyle\sin^2\theta_0}{\displaystyle\Delta_1^2}
+\frac{\displaystyle\pi|\sin2\theta_0|}{\displaystyle8\Delta_1\Delta_2}\right)
\enspace .
\label{thcylt}
\end{equation}
As earlier, the answer for polar phase and for $(1,i)$-phase of $E_{1g}$-type
of pairing is obtained from (\ref{thcylt}) in the limit
$\Delta_2\rightarrow\infty$.

For fairly low superfluid flow, when the condition $p_fv_s\ll
T\ll\Delta_{max}$ is satisfied, the expression for the thermal conductivity can
be obtained, substituting $I_{yy}= \frac{\displaystyle \pi\omega}{\displaystyle
8\Delta_1}$ together with Eq.(\ref{tau3p}) into Eq.(\ref{thc}):
\begin{equation}
\left.
\frac{\displaystyle\kappa_{yy}}{\displaystyle\kappa_N(T_c)}=
\frac{\displaystyle\kappa_{yy}}{\displaystyle\kappa_N(T_c)}\right\vert_{v_s=0}-
\frac{\displaystyle 9(p_fv_s\cos\theta_0)^2\left(\Delta_2+\Delta_1\right)}{
\displaystyle 8\pi^2T_c\Delta_1\left(2\Delta_2+\Delta_1\right)}
\int_0^{\infty}\frac{\displaystyle \omega d\omega}{\displaystyle
\cosh^2\left(\frac{\omega}{2}\right)}{\rm Re}G_0(T\omega,0)
\enspace .
\label{thcylsf}
\end{equation}

In the particular case of polar phase we obtain from (\ref{thcylsf}),
(\ref{reg0}) the following dependence of the thermal conductivity upon the low
enough superfluid flow:
$$
\frac{\displaystyle\kappa_{yy}}{\displaystyle\kappa_N(T_c)}=
\frac{\displaystyle10.36T^3}{\displaystyle T_c\Delta_1^2}
\left[\ln^2\left(\frac{\displaystyle 2\Delta_1}{\displaystyle T}\right)
-1.34\ln\left(\frac{\displaystyle 2\Delta_1}{\displaystyle T}\right)+2.95
\right]- \quad\quad\qquad\qquad\qquad\qquad\qquad\qquad
$$
\begin{equation}
\qquad\qquad\qquad\qquad\qquad\qquad\qquad\qquad\qquad -\frac{\displaystyle
0.38(p_fv_s\cos\theta_0)^2T}{ \displaystyle T_c\Delta_1^2}\left[
\ln\left(\frac{\displaystyle 2\Delta_1}{\displaystyle T}\right)-0.45\right]
\enspace .
\label{thcylsfpp}
\end{equation}

Spatial averaging of $\kappa_{yy}$ over that part of the intervortex space,
where Eq.(\ref{thcylsf}) is valid (~that is over its exterior part
$p_fv_s(r)\lesssim T$~), leads to the thermal conductivity, which diminishes
with increasing weak magnetic field for all three types of pairing according to
Eq.(\ref{thca}) (where one should change indices $z\to y$ and retain only sign
minus in front of $A_{yy}$). Disregarding the logarithmic terms, we find that
the function $A_{yy}(T)$ is roughly a linear function of temperature for each
type of pairing discussed. Furthermore, the interior part of the
intervortex space (~$T\lesssim p_fv_s(r)$~), where one can use
Eq.(\ref{thcylt}), results in $\kappa_{yy}$, which  rises with magnetic field
$\propto\frac{\displaystyle B}{\displaystyle B_{c2}}
\ln\left(\frac{\displaystyle B_{c2}}{\displaystyle B}\right)$. These two
regions compete with each other as it was above for $\kappa_{zz}$. But even in
the case $T\lesssim p_fv_s(r)$ large distances from vortex cores $\xi\ll
r\lesssim\xi \sqrt{B_{c2}/B}$ dominate here leading to additional logarithmic
factor, as compared to Eqs.(\ref{thcp1}), (\ref{thcp2}).  Possibly, this
logarithmic factor is responsible for exceeding in several times of
$\kappa_{yy}$ as compared to $\kappa_{zz}$, which is experimentally observed in
$UPt_3$ \cite{fl1,fl2}. For quantitative consideration of this problem, one
should evidently go beyond the approximation of locally quasiuniform superflow
in describing the mixed state.

\section{Scaling of the thermal conductivity at low fields and
low temperatures}

Thermodynamic and transport characteristics of superconductors with nodes may
exhibit scaling behavior at low fields and low temperatures.  This possibility
for high-temperature superconductors was studied theoretically
in\cite{kop1,kop2,plee,kop3}. In particular, it was shown, that the electronic
thermal conductivity of a two-dimensional $d$-wave superconductor with four
lines of nodes on a cylindrical Fermi surface may be represented as
$\kappa_{ij}\sim TF_{ij}(\alpha T/B^{1/2})$\cite{plee}. One of the important
assumptions underlying this result is that contributions to the thermal
conductivity from quasiparticles in the intervortex space far enough from
vortex cores dominate.  In contrast to the thermodynamic quantities, for
transport phenomena scattering processes are of importance even in clean limit.
We note, that properties of disorder impurity potentials considered in Ref.\
\onlinecite{plee} may be suitable for the Born scatterers, not in the unitary
limit.

Superconducting $UPt_3$ essentially differs from the superconductors just
mentioned above. It is three-dimensional hexagonal superconductor with the
order parameter, which is believed to have both line of nodes on the equator of
the Fermi surface and point nodes on its poles.  While the strength of impurity
scattering for high-temperature superconductors is not yet definitely
determined, for heavy-fermion superconductors (in particular, for $UPt_3$)
there are various experimental results and the physical
background\cite{pet3,varma} indicating to the impurity scattering very close
to the unitarity limit. Thus, the problem of scaling behavior of the thermal
conductivity in superconductors like $UPt_3$ still has not been theoretically
studied properly.

At the same time the behavior quite close to the scaling one has been
experimentally established recently for the thermal conductivity at low fields
and low temperatures for superconducting $UPt_3$ both for the component
$\kappa_{zz}$ under the magnetic field parallel to z-axis and for $\kappa_{yy}$
in the case of the field applied along $y$-axis\cite{fl2}. These are just the
cases which we consider in the article. Two forms of scaling are of interest
for discussion of the experimental data of Ref.\ \onlinecite{fl2}:
\begin{equation}
\kappa_{ii}= T^3F_{ii}(x) \enspace , \hspace{1cm}
x=\frac{\displaystyle T}{\displaystyle T_c}
\sqrt{\frac{\displaystyle B_{c2}}{\displaystyle B}} \enspace ;
\label{sc1}
\end{equation}
\begin{equation}
\frac{\displaystyle\kappa_{ii}(T,B)}{\displaystyle\kappa_{ii}(T,B=0)}=
g_{ii}\left(x\right) \enspace .
\label{sc2}
\end{equation}

In order to check whether our results are in agreement with any of these
scaling behaviors, let us specify firstly power law dependences of the
thermal conductivity upon temperature (under the condition $\gamma\lesssim
T\ll\Delta_{max}$) for three particular types of pairing in question in the
absence of the magnetic field.

According to Eq.(\ref{tauun1}), in order to estimate frequency dependence of
the relaxation time in the unitarity limit, one should find the corresponding
behavior of ${\rm Im}G_0$, ${\rm Re}G_0$.  Contributions to ${\rm Im}G_0$,
${\rm Re}G_0$ from the line of nodes in the equatorial plane (subscript $l$),
linear and quadratic point nodes at the poles of the Fermi surface (subscripts
$p1$ and $p2$ respectively) are as follows
\begin{equation}
\begin{array}{lll}
\left\{
\begin{array}{l}
{\rm Im}G_{0,l}(\omega)=\frac{\displaystyle\pi\omega}{\displaystyle2\Delta_1}\\
{\rm Im}G_{0,p1}(\omega)=\frac{\displaystyle\omega^2}{
\displaystyle\Delta_2^2}\\
{\rm Im}G_{0,p2}(\omega)=\frac{\displaystyle\pi\omega}{
\displaystyle 4\Delta_2}
 \\
\end{array}
\right.
&\begin{array}{l}\\ \\ \enspace , \end{array}&
\hspace{1cm} \left\{
\begin{array}{l}
{\rm Re}G_{0,l}(\omega)\approx\frac{\displaystyle\omega}{\displaystyle
\Delta_1}\ln\left(\frac{\displaystyle A_l\Delta_1}{\displaystyle
\omega}\right)\\
{\rm Re}G_{0,p1}(\omega)\approx\frac{\displaystyle
A_{p1}\omega}{\displaystyle\Delta_2}\\
{\rm Re}G_{0,p2}(\omega)\approx\frac{\displaystyle\omega}{
\displaystyle \Delta_2}\ln\left(\frac{\displaystyle A_{p2}\Delta_2}{
\displaystyle\omega}\right) \enspace .\\
\end{array}
\right.
\\
\end{array}
\label{g0}
\end{equation}
One should emphasize that for all three cases we consider, the behavior of
${\rm Re}G_0$ (in contrast to ${\rm Im}G_0$) is governed, even at low
frequencies, by the behavior of the order parameter not only near the nodes,
but over the whole Fermi surface. So, the given expressions for ${\rm Re}G_0$
are approximate ones and constants $A_{l}$, $A_{p1}$, $A_{p2}$ can't be
determined unambiguously unless the behavior of the order parameter all over
the Fermi surface is known.  For the polar state $\Delta=\Delta_1\cos\theta$
and a spherical Fermi surface we get $A_l=2$ from the comparison of
(\ref{reg0}) and (\ref{g0}) at $\omega\ll\Delta_1$. Let the order parameter in
more general case have the form $|\Delta|=\Delta_0f(x)$, where $x=\cos\theta$
and $f(x)\approx f'(0)x$ for $|x|\ll 1$, $|f(x)|\approx |f'(1)|(1\mp x)$ for
$(1\mp x)\ll 1$.  Then one can show for sufficiently small frequencies
\begin{equation}
{\rm Re}G_0(\omega)=\zeta\frac{\displaystyle \omega}{\displaystyle \Delta_0}
\ln\left(\frac{\displaystyle A\Delta_0}{\displaystyle\omega}\right) \enspace ,
\qquad
{\rm Im}G_0(\omega)=\zeta\frac{\displaystyle\pi\omega}{\displaystyle
2\Delta_0}\enspace ,
\end{equation}
where
\begin{equation}
\zeta=\frac{\displaystyle 1}{\displaystyle|f'(0)|}+\frac{\displaystyle 1}{
\displaystyle |f'(1)|}, \qquad
\zeta\ln A=\lim_{\alpha\to 0}\left(\int_\alpha^{1-\alpha}\frac{\displaystyle
dx}{\displaystyle f(x)}+\frac{\displaystyle \ln(2\alpha|f'(0)|)}{\displaystyle
|f'(0)|}+\frac{\displaystyle\ln(2\alpha|f'(1)|) }{\displaystyle |f'(1)|}
\right).
\end{equation}
So, in calculating $A$ one has to integrate $1/f(x)$ over the whole Fermi
surface.

On the other side, if one does not fix from the very beginning the behavior of
the order parameter all over the Fermi surface but only near the nodes,
constants of this origin may be considered as fitting parameters in comparison
of the theoretical results with experimental data under the corresponding
conditions. These fitting parameters, generally speaking, may manifest weak
temperature dependence in the low temperature region, associated with the
respective low-temperature dependence of the order parameter. Any particular
choice of the parameters corresponds yet with a large number of basis functions
of the representation, rather than only with a unique particular one.

One can easily see from Eqs.(\ref{g0}), (\ref{tauun1}), that if only
one kind of nodes is present, then at low frequencies
\begin{equation}
\begin{array}{c}
\frac{\displaystyle\tau_{s,l}(\omega)}{\displaystyle \tau_N}\approx
\frac{\displaystyle 2\omega}{\displaystyle\pi\Delta_1}
\left[\frac{\displaystyle\pi^2 }{\displaystyle4}+\ln^2\left(
\frac{\displaystyle \omega}{\displaystyle A_l\Delta_1}\right)\right]\, ,
\hspace{1cm}
\frac{\displaystyle\tau_{s,p1}(\omega)}{\displaystyle\tau_N}\approx
A^2_{p1}+\frac{\displaystyle\omega^2}{\displaystyle\Delta_2^2}\, ,\\
\frac{\displaystyle\tau_{s,p2}(\omega)}{\displaystyle\tau_N}\approx
\frac{\displaystyle 4\omega}{\displaystyle\pi
\Delta_2}\left[\frac{\displaystyle\pi^2 }{\displaystyle16}+\ln^2\left(
\frac{\displaystyle \omega}{\displaystyle A_{p2}\Delta_2}\right)\right]\, .\\
\end{array}
\label{tlp}
\end{equation}
Low temperature dependence of the relaxation time of the form
$T\ln^2(\Delta/T)$, found for the polar state in\cite{pet3}, is in agreement
with the expression for $\tau_{s,l}$.

For $(1,i)$-phases of $E_{1g}$- and $E_{2u}$-types of pairing with hybrid gap
functions, having both the line of nodes and point nodes, the low-frequency
dependences of the relaxation times are similar to the case of the polar state.
Although the particular form of the order parameter all over the Fermi surface,
for instance, point nodes result in the change of constants in the approximate
expression for $\tau_{s,l}$, which then differ for each type of pairing. So, we
write
\begin{equation}
\frac{\displaystyle\tau_{s,1g}(\omega)}{\displaystyle\tau_N}\approx
\frac{\displaystyle2\omega}{\displaystyle\pi\Delta_1}\left[\ln^2\left(\frac{
\displaystyle A_{1g}\omega}{\displaystyle\Delta_1}\right)+
\frac{\displaystyle\pi^2}{\displaystyle 4}\right]
\enspace ,  \qquad\qquad
\\
\frac{\displaystyle\tau_{s,2u}(\omega)}{\displaystyle\tau_N}\approx
\frac{\displaystyle2\omega}{\displaystyle\pi\Delta_{ef}}\left[\ln^2\left(\frac{
\displaystyle A_{2u}\omega}{\displaystyle\Delta_{ef}}\right)+
\frac{\displaystyle\pi^2}{\displaystyle 4}\right]
\enspace ,
\label{taugu}
\end{equation}
where\, $\Delta_{ef}=2\Delta_1\Delta_2/(\Delta_1+2\Delta_2)$.

Contributions from the line and point nodes to low-frequency behaviors of
$I_{ij}(\omega)$ are
\begin{equation}
\begin{array}{lll}
\left\{
\begin{array}{l}
I_{zz,l}(\omega)=\frac{\displaystyle\pi\omega^3}{\displaystyle16\Delta_1^3}\\
I_{zz,p1}(\omega)=\frac{\displaystyle\omega^2}{\displaystyle
3\Delta_2^2} \\
I_{zz,p2}(\omega)=\frac{\displaystyle\pi\omega}{\displaystyle
8\Delta_2}
\\
\end{array}
\right.
&\begin{array}{l}\\ \\ \enspace , \end{array}&
\hspace{1cm} \left\{
\begin{array}{l}
I_{yy,l}(\omega)=\frac{\displaystyle\pi\omega}{\displaystyle8\Delta_1}\\
I_{yy,p1}(\omega)=\frac{\displaystyle\omega^4}{\displaystyle
30\Delta_2^4} \\
I_{yy,p2}(\omega)=\frac{\displaystyle\omega^2}{\displaystyle
24\Delta_2^2}
\enspace .\\
\end{array}
\right.   \\
\end{array}
\label{iii}
\end{equation}

Substituting Eqs.(\ref{tlp}) -- (\ref{iii}) into (\ref{thc}), we obtain the
following leading low-temperature terms for $\kappa_{zz}$ under the condition
$\gamma\lesssim T\ll T_c$:
\begin{equation}
\left\{\begin{array}{l}
\frac{\displaystyle\kappa_{zz,p1}}{\displaystyle\kappa_N(T_c)}\approx
\frac{\displaystyle 13.8 T^3 A_{p1}^2}{\displaystyle T_c\Delta_2^2}\\
\frac{\displaystyle\kappa_{zz,p2}}{\displaystyle\kappa_N(T_c)}\approx
\frac{\displaystyle 20.7T^3}{\displaystyle T_c\Delta_2^2}
\left[\ln^2\left(\frac{\displaystyle A_{p2}\Delta_2}{\displaystyle
4.67T}\right)+0.82\right]\\
\frac{\displaystyle\kappa_{zz,\phantom{1}l}}{\displaystyle\kappa_N(T_c)}\approx
\frac{\displaystyle 161.77 T^5}{\displaystyle  T_c\Delta_1^4}\left[
\ln^2\left(\frac{\displaystyle A_l\Delta_1}{\displaystyle 6.59 T}\right)+
2.61\right]\\
\frac{\displaystyle\kappa_{zz,1g}}{\displaystyle\kappa_N(T_c)}\approx
\frac{\displaystyle 45.14 T^4}{\displaystyle T_c\Delta_2^2\Delta_1}\left[
\ln^2\left(\frac{\displaystyle \Delta_1}{\displaystyle 5.61 A_{1g}T}\right)
+2.64\right]\\
\frac{\displaystyle\kappa_{zz,2u}}{\displaystyle\kappa_N(T_c)}\approx
\frac{\displaystyle 10.36 T^3}{\displaystyle T_c\Delta_{ef}\Delta_2}\left[
\ln^2\left(\frac{\displaystyle \Delta_{ef}}{\displaystyle 4.67A_{2u}T}\right)
+2.67\right]\\
\end{array}
\right.
\label{zz}
\end{equation}

and for $\kappa_{yy}$:
\begin{equation}
\left\{\begin{array}{l}
\frac{\displaystyle\kappa_{yy,p1}}{\displaystyle\kappa_N(T_c)}\approx
\frac{\displaystyle 43.1 T^5 A_{p1}^2}{\displaystyle T_c\Delta_2^4}\\
\frac{\displaystyle\kappa_{yy,p2}}{\displaystyle\kappa_N(T_c)}\approx
\frac{\displaystyle 11.3 T^4}{\displaystyle T_c\Delta_2^3}
\left[\ln^2\left(\frac{\displaystyle A_{p2}\Delta_2}{\displaystyle
5.61 T}\right)+0.78\right]\\
\frac{\displaystyle\kappa_{yy,\phantom{1}l}}{\displaystyle\kappa_N(T_c)}\approx
\frac{\displaystyle 10.36 T^3}{\displaystyle  T_c\Delta_1^2}\left[
\ln^2\left(\frac{\displaystyle A_l\Delta_1}{\displaystyle 4.67 T}\right)+
2.67\right]\\
\frac{\displaystyle\kappa_{yy,1g}}{\displaystyle\kappa_N(T_c)}\approx
\frac{\displaystyle 10.36 T^3}{\displaystyle T_c\Delta_1^2}\left[
\ln^2\left(\frac{\displaystyle \Delta_1}{\displaystyle 4.67 A_{1g}T}\right)
+2.67\right]\\
\frac{\displaystyle\kappa_{yy,2u}}{\displaystyle\kappa_N(T_c)}\approx
\frac{\displaystyle 10.36 T^3}{\displaystyle T_c\Delta_{ef}\Delta_1}\left[
\ln^2\left(\frac{\displaystyle \Delta_{ef}}{\displaystyle 4.67A_{2u}T}\right)
+2.67\right]
\enspace .  \\
\end{array}
\right.
\label{yy}
\end{equation}

The first three results in Eq.(\ref{zz}) (as well as in Eq.(\ref{yy})) are in
agreement with those obtained in\cite{bar2}, where numerical coefficients were
not specified. They concern the cases when only one kind of nodes is presented
(both the line of nodes or one kind of point nodes, but not a hybrid gap
function). The results in Eqs.(\ref{zz}), (\ref{yy}) concerning hybrid gap
functions, are new. The particular example of $\kappa_{zz,1g}$-component of the
thermal conductivity demonstrates, that in the unitary limit in the presence of
several kinds of nodes (that is for a hybrid gap function) the index of power
law behavior of the low-temperature thermal conductivity may be greater than
the least index among those taking place for superconductors with one separate
kind of the nodes (~in particular, in discussing of the $E_{1g}$-case -- for
superconductors with the linear point nodes~).

Since the same relaxation time enters the expressions for both $\kappa_{zz}$
and $\kappa_{yy}$, then for each particular type of pairing there are relations
between the coefficients in those expressions, which govern the behavior of the
anisotropy ratio $\kappa_{zz}/\kappa_{yy}$.  For instance, as for $E_{2u}$ type
of pairing the both quantities $I_{zz,2u}$ and $I_{yy,2u}$ are proportional to
$\omega$, the anisotropy ratio of the thermal conductivity in leading
approximation is determined simply by the ratio $I_{zz,2u}/I_{yy,2u}=
\Delta_1(T)/\Delta_2(T)$, that is only weakly depends upon temperature.

For $E_{1g}$ case we have $I_{zz,1g}\propto \omega^2$, while
$I_{yy,1g}\propto\omega$. As a consequence, the anisotropy ratio of the thermal
conductivity for $E_{1g}$ type of pairing essentially depends upon temperature:
\begin{equation}
\frac{\displaystyle \kappa_{zz,1g}}{\displaystyle \kappa_{yy,1g}}=\frac{
\displaystyle 4.36 T\Delta_1}{\displaystyle \Delta_2^2}\left\{1-
\frac{\displaystyle0.37 \ln\left(\frac{\displaystyle\Delta_1}{\displaystyle
4.67 A_{1g}T}\right)}{\displaystyle
\ln^2\left(\frac{\displaystyle \Delta_1}{\displaystyle 4.67 A_{1g}T}\right)
+2.67}\right\}
\enspace .
\label{rat1g}
\end{equation}
Essential temperature dependence of the ratio for $E_{1g}$ model was noticed
earlier on the basis of numerical results in\cite{nor1,hir3}.

The analysis of experimental data of Ref.\ \onlinecite{fl1} on the anisotropy
ratio of the thermal conductivity in $UPt_3$ at low temperatures
$\gamma\lesssim T\ll T_c$ seems to permit one to distinguish between $E_{2u}$
and $E_{1g}$ representations in favor of $(1,i)$-phase of $E_{2u}$-type of
pairing. As it follows from\cite{fl1} for their particular clean samples, the
temperature interval for which the low-temperature power law behavior of the
thermal conductivity takes place is approximately $0.07<T/T_{c-}<0.15$.
According to Eq.(\ref{rat1g}), the anisotropy ratio should increase in more
than $1.78$ times, when the temperature changes from $T=0.07T_{c-}$ to
$T=0.14T_{c-}$. This seems to be in a contradiction with the experiment (see,
Fig.\ 2 in\cite{fl1}), which shows the increase of the anisotropy ratio only on
$10-12$ per cent with the temperature change discussed. In other words, the
fitting parameters available for the $E_{1g}$ case in Eqs.\ (\ref{zz}),
(\ref{yy}) allow to describe properly experimental results for the given
temperature interval both for $\kappa_{zz}$ or for $\kappa_{yy}$, and not for
both components simultaneously.

While $E_{1g}$ type of pairing leads to noticeable overestimation of the
low-temperature dependence of the anisotropy ratio $\kappa_{zz}/\kappa_{yy}$ in
$UPt_3$, within the framework of $E_{2u}$ type of pairing we get, for the first
sight, the underestimation of that dependence. Although for superconductors
with nodes low-temperature deviation of the order parameter from its
zero-temperature value is not exponentially small (as for $s$-wave isotropic
case), but manifests power-law dependences, the temperature dependence of the
quantity $\Delta_1(T)/\Delta_2(T)$ is yet too weak in order to explain the
observed change of the ratio on $10-12$ per cent. However, the temperature
dependence discussed can be described if one keeps in the expression for the
anisotropy ratio, obtained within $E_{2u}$ type of pairing, the first
low-temperature correction to the leading term. For this purpose one should
specify next terms in the expansion of the momentum dependence of the order
parameter near nodes. We let $|\Delta|\approx\Delta_1|(\pi/2-\theta)+
L(\pi/2-\theta)^3|$ for $|\theta-(\pi/2)|\ll 1$, and $|\Delta|\approx \Delta_2(
\theta^2-D\theta^4)$ for $|\theta|\ll 1$. Then the calculations of the
anisotropy ratio under the condition $\gamma\lesssim T\ll T_{c}$ result in
\begin{equation}
\frac{\displaystyle\kappa_{zz,2u}}{\displaystyle \kappa_{yy,2u}}=
\frac{\displaystyle\Delta_1}{\displaystyle\Delta_2}\left(1+
4.36\left(D-0.58-0.12\frac{\displaystyle\Delta_1}{\displaystyle\Delta_2}\right)
\frac{\displaystyle T}{\displaystyle\Delta_2}S(T)\right)
\enspace ,
\label{rat2u}
\end{equation}
where the quantity
\begin{equation}
S(T)=\frac{\displaystyle \ln^2(\Delta_{ef}/(5.61A_{2u}T))+2.64}{\displaystyle
\ln^2(\Delta_{ef}/(4.67A_{2u}T))+2.67}
\label{s}
\end{equation}
is close to unity.

We see, that this relation is sufficiently sensitive to the value of the
coefficient $D$, while the effect of $L$ is beyond the first correction to the
leading term.  Taking $D$ as a fitting parameter, one can easily describe the
observed change of the anisotropy ratio for $0.07<T/T_{c-}<0.15$. For example,
let $\Delta_2\approx 3\Delta_1\approx 12T_{c-}$, then we find $D\approx 5.2$.
Particular values for coefficients are obtained here for a spherical Fermi
surface and they, of course, depend upon the form of the Fermi surface.  It is
of importance to emphasize, however, that the anisotropy of the Fermi surface
doesn't change our qualitative conclusion itself. It is based on the fact, that
$\kappa_{zz,1g}/\kappa_{yy,1g}$ is roughly proportional to the temperature,
while $\kappa_{zz,2u}/\kappa_{yy,2u}$ weakly depends on $T$, being at the same
time sufficiently sensitive to the coefficient $D$ in the next term of the
expansion of momentum dependence of the order parameter near point nodes. Note
relatively large values of $D$, to which our analysis results in.  The
necessity for such a value of $D$ should be taken into account when the
particular basis function is chosen for the numerical study.  There is, of
course, the problem of describing the temperature dependence of the anisotropy
ratio at ultra low temperatures $T\lesssim \gamma$, which should take account
of steeper slope of the curve, which the experiments show for
$\kappa_{zz}/\kappa_{yy}$ in the crossover between two regimes. This problem is
not considered here.

For higher temperatures the behavior of the anisotropy ratio becomes much more
sensitive to the particular form of the order parameter all over the Fermi
surface, not to its behavior mostly near nodes. As a consequence, the problem
of discrimination between types of pairing in considering the anisotropy ratio
of the thermal conductivity is essentially ambiguous in that case\cite{nor1}.

Furthermore, it follows from Eqs.\ (\ref{thcp1}), (\ref{thcp2}), (\ref{thca}),
(\ref{zz}), (\ref{yy}) (~in particular, from temperature dependences of
$A_{zz}(T)$, $A_{yy}(T)$~), that for temperatures in question scaling behavior
(\ref{sc1}) (~as well as (\ref{sc2})~) is valid (in disregarding logarithmic
terms) for both components $\kappa_{ii}(T,B)$ ($i=x,y$) and for both limiting
cases $\gamma\lesssim T\lesssim T_c\sqrt{ B/B_{c2}}$,\ \ $T_c\sqrt{B/B_{c2}}\ll
T\ll T_c$ just for $E_{2u}$-representation. For $E_{1g}$ type of pairing
deviations from scaling are more noticeable for $\kappa_{zz}$-component both in
the temperature dominating region $T_c\sqrt{B/B_{c2}}\ll T\ll T_c$ (if one uses
the form (\ref{sc1})), or in the field dominating region $\gamma\lesssim
T\lesssim T_c\sqrt{ B/B_{c2}}$ (for the form (\ref{sc2})). In order to
distinguish between two models, considering the magnetic field dependence of
$\kappa$, one could introduce, along with the zero-field anisotropy ratio
discussed just now, the other characteristic as well. It is
the anisotropy ratio of the form $\left(\kappa_{zz}(T,B)-\kappa_{zz}(T,B=0)
\right)/\left(\kappa_{yy}(T,B)-\kappa_{yy}(T,B=0)\right)=A_{zz}(T)/A_{yy}(T)$
under the temperature dominating condition. Analogously to the anisotropy ratio
in zero field, quantity $A_{zz}(T)/A_{yy}(T)$, in accordance with
Eqs.(\ref{thce}), (\ref{thcylsf}), (\ref{g0}), is only weakly temperature
dependent function within the framework of $E_{2u}$ model, while it is roughly
proportional to $T$ for the $E_{1g}$ type of pairing. The anisotropy of the
Fermi surface doesn't change the conclusion. Unfortunately, for $UPt_3$
under the field $B>B_{c1}$ low temperatures discussed do not belong to the
temperature dominating region $T_c\sqrt{B/B_{c2}}\ll T\ll T_c$.

The other test for the discrimination between possible types of pairing in
superconducting $UPt_3$, which is based on the presence of superflow, is as
follows.  Since the only important qualitative difference between order
parameters of $E_{2u}$ and $E_{1g}$ types of pairing is the multiplicity of
point nodes on the poles of the Fermi surface, it looks reasonable, for
maximizing the difference between the effects, to consider $\kappa_{zz}$
component of thermal conductivity under the presence of the supercurrent along
the $z$-axis. In order not to mix various directions of the superflow in this
case, let it be the uniform transport supercurrent in the absence of the
magnetic field. Such a problem is associated with possible experiments on thin
films or whiskers\cite{sud}. One can use for both types of pairing
Eq.(\ref{tau3p}) for the relaxation time, inserting there $\theta_0=0$.
Further, under the given conditions
\begin{equation}
I_{zz,2u}(\omega, v_s)=
\left\{\begin{array}{ccl}
\frac{\displaystyle \pi\omega}{\displaystyle 8\Delta_2}& , & p_fv_s\ll \omega
\enspace , \\
&&\\
\frac{\displaystyle \pi p_fv_s}{\displaystyle 8\Delta_2}
& , &   p_fv_s\gg \omega \enspace ; \\
\end{array}
\right.
\quad
I_{zz,1g}(\omega, v_s)=
\left\{\begin{array}{ccl}
\frac{\displaystyle \omega^2+(p_fv_s)^2}{\displaystyle 3\Delta_2^2}& , &
p_fv_s\ll \omega \enspace ,\\
&&\\
\frac{\displaystyle (p_fv_s)^2}{\displaystyle 3\Delta_2^2}
& , &   p_fv_s\gg \omega \enspace . \\
\end{array}
\right.
\label{wh}
\end{equation}

Inserting these results into (\ref{thc}), we obtain under the temperature
dominating conditions $p_fv_s\ll T\ll\Delta_{max}$:
\begin{equation}
\left.
\frac{\displaystyle\kappa_{zz,2u}}{\displaystyle\kappa_N(T_c)}=
\frac{\displaystyle\kappa_{zz,2u}}{\displaystyle\kappa_N(T_c)}\right\vert_{v_s=0}-
\frac{\displaystyle 9(p_fv_s)^2\left(\Delta_2+\Delta_1\right)}{
\displaystyle 8\pi^2T_c\Delta_2\left(2\Delta_2+\Delta_1\right)}
\int_0^{\infty}\frac{\displaystyle \omega d\omega}{\displaystyle
\cosh^2\left(\frac{\omega}{2}\right)}{\rm Re}G_0(T\omega,0)
\enspace ,
\end{equation}
$$
\left.
\frac{\displaystyle\kappa_{zz,1g}}{\displaystyle\kappa_N(T_c)}=
\frac{\displaystyle\kappa_{zz,1g}}{\displaystyle\kappa_N(T_c)}
\right\vert_{v_s=0}+
\quad\qquad\qquad\qquad\qquad\qquad\qquad\qquad\qquad\qquad\qquad\qquad\qquad\qquad\qquad
$$
\begin{equation}
\qquad\qquad
\frac{3}{\pi^2}\frac{T^2}{\Delta_1 T_c}\frac{(p_fv_s)^2}{\Delta_2^2}
\int_0^{\infty}\frac{\displaystyle \omega d\omega}{\displaystyle
\cosh^2\left(\frac{\omega}{2}\right)}
\left[\left({\rm Re}G_0(T\omega,0)\frac{\Delta_1}{T}-\frac{\omega}{4}\right)^2+
\frac{\omega^2}{4}\left(\pi^2-\frac{1}{4}\right)\right]
\enspace ,
\end{equation}
and for the superflow dominating regime $T\ll p_fv_s\ll \Delta_{max}$:
\begin{equation}
\frac{\displaystyle\kappa_{zz,2u}}{\displaystyle\kappa_N(T_c)}=
\frac{3\pi^2}{32}\frac{T}{T_c}\frac{(p_fv_s)^2}{\Delta^2_2}
\enspace , \qquad
\frac{\displaystyle\kappa_{zz,1g}}{\displaystyle\kappa_N(T_c)}=
A_{p1}^2\frac{T}{T_c}\frac{(p_fv_s)^2}{\Delta^2_2}
\enspace .
\end{equation}
We see for the given direction of the supercurrent, that $\kappa_{zz}$  is the
nonmonotonous function upon the value of the supercurrent only in the case of
$E_{2u}$ type of pairing, while for the $E_{1g}$ model the thermal conductivity
at low temperatures
monotonously changes with the superflow velocity (see Fig.\ 3).  This kind of
experiments seems to be quite useful for the discrimination between candidates
for the type of pairing in superconducting $UPt_3$.

One of the characteristic features of clean superconductors with nodes is the
strong low energy dependence of the relaxation time, both in the Born
approximation and in the unitary limit\cite{rice,pet3,pet1,pet2}(~see as well
Eqs.(\ref{tau1}), (\ref{pol1tau}), (\ref{tau3p}), (\ref{rtpp}) in the above~).
This is essential for finding the low temperature behavior of the thermal
conductivity in clean limit. It is worth noting that in the unitary limit the
relaxation time may manifest weak dependence upon energy for
$\omega/\Delta_{max}>0.1$, so that for not very clean samples with
$0.1\Delta_{max}\lesssim \gamma$ a model with energy independent relaxation
time may present a reasonable approximation\cite{varma,pet1}. For these values
of $\gamma$ the conditions $\gamma\lesssim T\ll T_c$, we are interested in in
the article, are not satisfied. However, the low temperature region
$\gamma\lesssim T\ll T_c$ does exist for samples of Ref.\ \onlinecite{fl1,fl2},
for which their authors give the estimate $\gamma\approx 0.017K$ (one gets for
this value $\gamma\approx 0.038T_{c-}\sim 0.01\Delta_{max}$).

We note, that in the presence of scattering by impurities, which is close to
the unitarity limit, both the relaxation time and the electronic thermal
conductivity of two-dimensional sufficiently clean d-wave superconductors would
manifest analogous nonmonotonous dependences upon the magnetic field, which are
found above for the three-dimensional superconductors with nodes.

\section{Thermal conductivity dominated by impurity-induced bound states }

Impurity-induced bound states dominate the thermal conductivity  under the
condition $T, v_f{\rm v}_s\lesssim\gamma$. It is essential, that in this limit
results turn out to be independent of the relation between $T$ and $v_f{\rm
v}_s$. The spatial averaging over the intervortex space would contain both the
region where $v_f{\rm v}_s\lesssim\gamma$ and a part of the intervortex space
where the opposite condition is valid (~since $v_f{\rm v}_s(\xi)\sim T_{c}$~).
Hence impurity-induced bound states could influence essentially the spatially
averaged thermal conductivity only for large enough values of $\gamma$.  So,
for clean superconductors with sufficiently small $\gamma$ one can disregard
the influence of impurity bound states on the magnetic field dependence of the
thermal conductivity.  Nevertheless, it is of interest to consider how the
uniform superfluid flow influences the thermal conductivity dominated by
impurity induced bound states. Below we show, that in the unitarity limit the
thermal conductivity dominated by impurity induced bound states diminishes with
increasing superflow field under various conditions. Hence, there is no
nonmonotonies in the superflow field dependence of the thermal conductivity at
least unless the relation $T, v_f{\rm v}_s\lesssim\gamma$ is broken.

Excitation energy renormalized by impurities satisfies the relation
$\tilde\omega(-\omega)=-\tilde\omega^*( \omega)$\cite{graf1,rai1}. Thus,
retaining for sufficiently small $\omega$ a pair of terms in the expansion of
$\tilde\omega(\omega)$ in powers of $\omega$, we have:
\begin{equation}
\tilde\omega(\omega)\simeq i\gamma +a\omega \enspace ,
\end{equation}
where $\gamma$ and $a$ are real positive parameters. Parameter $a$ is a
function of $\gamma$ and, in particular, in the unitary limit in the absence of
the condensate flow field it takes the form
\begin{equation}
a=\frac{\displaystyle \left<\left(\gamma^2+|\tilde\Delta|^2\right)^{-1/2}
\right>_{S_f}}{\displaystyle\left<\left(\gamma^2+
2|\tilde\Delta|^2\right)\left(\gamma^2+|\tilde\Delta|^2\right)^{-3/2}
\right>_{S_f}}\enspace .
\end{equation}
For sufficiently small $\gamma$ one gets $a=1/2$.
In the presence of the superflow field parameter $\gamma$ is a function of
${\bf v}_s$, and up to the first correction to its zero field value we get
\begin{equation}
\gamma\simeq\gamma_0 + b{\rm v}^2_s \enspace .
\label{gvs}
\end{equation}

To estimate $b$ we make use of the results of Ref.\ \onlinecite{bar1}, and
obtain that in the unitary limit this coefficient is negative and $|b|\sim
v_f^2/\gamma_0 $, at least within the logarithmic accuracy.

We use the expansion of the integrand in Eq.(\ref{kappaij}) over a parameter
$(a\omega- \bbox{v}_f{\bf v}_s)\gamma/(\gamma^2+|\tilde\Delta|^2)$, which is
supposed to be small. Renormalized order parameter can be taken in this limit
for $\omega=0$, since we are interested in most important terms which are
linear in temperature. Retaining the corrections to the zero field term up to
second order in the superflow field, we obtain
$$
\kappa_{ii}= \frac{\displaystyle \pi^2 N_fT}{\displaystyle
3}\left <\enspace v^2_{f,i}\left[
\frac{\displaystyle \gamma_0^2}{
\displaystyle (\gamma_0^2+|\tilde\Delta|^2)^{3/2}}+
\left(\bbox{v}_f{\bf v}_s\right)^2\left(
\frac{\displaystyle \gamma_0^2}{
\displaystyle (\gamma_0^2+|\tilde\Delta|^2)^{5/2}}
-\frac{\displaystyle 5\gamma_0^4}{
\displaystyle 2(\gamma_0^2+|\tilde\Delta|^2)^{7/2}}\right)
\right.
\right.
\quad\quad
$$
\begin{equation}
\qquad\qquad\qquad\qquad\qquad\qquad\qquad\qquad
\left.
\left.
+b{\rm v}_s^2\left(\frac{\displaystyle 2\gamma_0}{
\displaystyle (\gamma_0^2+|\tilde\Delta|^2)^{3/2}}-
\frac{\displaystyle 3\gamma_0^3}{
\displaystyle (\gamma_0^2+|\tilde\Delta|^2)^{5/2}}
\right)\right]\right>_{S_f} \enspace .
\label{kapg}
\end{equation}

In the presence of nodes of the order parameter only narrow regions on the
Fermi surface near those nodes are of importance in Eq.(\ref{kapg}) under the
condition $\gamma\ll\Delta_{max}$. This permits to consider contributions from
each kind of nodes separately. Due to the same reason corrections containing
${\bf v}_s$ and coming from the corresponding dependence of $|\tilde\Delta|$ in
the first term in Eq.(\ref{kapg}), may be disregarded in this case, since they
would have an additional small factor $\sim(\gamma_0/\Delta_{max})^2$
associated with small characteristic angular regions near the nodes.

Let the order parameter have the line of nodes on the equator of a spherical
Fermi surface:  $|\Delta|=\Delta_1|\frac{\pi}{2}-\theta|$. Then we get from
Eq.(\ref{kapg}) the contribution from this line to the thermal conductivity
$\kappa_{yy}$ in the presence of the superflow, corresponding to the magnetic
field along $y$-axis:
\begin{equation}
\kappa_{yy}=\frac{\displaystyle \pi^2N_fv_f^2T}{\displaystyle 6\Delta_1}\left(
1-\frac{\displaystyle {\rm v}_{s,x}^2v_f^2}{\displaystyle 6\gamma_0^2}\right)
\enspace .
\label{kaply}
\end{equation}

Contributions from point nodes on the poles (~both linear and quadratic~) are
negligibly small for $\kappa_{yy}$. It is not the case for $\kappa_{zz}$.  In
the presence of magnetic field along $z$-axis we find from Eq.(\ref{kapg})
(within the logarithmic approximation) the following contribution from the line
of nodes for this component of the thermal conductivity:
\begin{equation}
\kappa_{zz}\simeq\frac{\displaystyle \pi^2\gamma_0^2N_fv_f^2T}{\displaystyle
3\Delta_1^3}\left[\ln\left(
\frac{\displaystyle 2A\Delta_1}{\displaystyle e\gamma_0}\right)  +
\frac{\displaystyle 2b{\rm v}_s^2}{\displaystyle \gamma_0}\ln\left(
\frac{\displaystyle 2A\Delta_1}{\displaystyle e^{3/2}\gamma_0}\right)\right]
\, .
\label{kaplz}
\end{equation}

At the same time the contribution from the linear point node is
\begin{equation}
\kappa_{zz}\simeq\frac{\displaystyle \pi^2 N_fv_f^2T}{\displaystyle
3}\left[
\frac{\displaystyle \gamma_0}{\displaystyle\Delta_2^2 }  +
\frac{\displaystyle b{\rm v}_s^2}{\displaystyle \Delta_2^2}\right]
\, .
\label{kapp1z}
\end{equation}
Making use of the estimation made above for the coefficient $b$, we have
disregarded here the term, which is in $(\gamma_0/\Delta_{2})^2$ times less
than the last term in (\ref{kapp1z}).

Furthermore, for quadratic point node one gets
\begin{equation}
\kappa_{zz}\simeq\frac{\displaystyle \pi^2 N_fv_f^2T}{\displaystyle
3}\left[
\frac{\displaystyle 1}{\displaystyle 2\Delta_2}  -
\frac{\displaystyle {\rm v}_s^2v_f^2}{\displaystyle24\gamma_0\Delta_2^2 }
\right]
\, .
\label{kapp2z}
\end{equation}

In the absence of the superfluid flow field results (\ref{kaply}) -
(\ref{kapp2z}) coincide with the ones obtained in \cite{graf1,graf2}.
According to Eqs.(\ref{kaply}) and (\ref{kapp2z}), thermal conductivity
diminishes with increasing superflow under the corresponding conditions.
Negative value of $b$ ensures the same qualitative conclusion for
Eqs.(\ref{kaplz}), (\ref{kapp1z}). Scaling relation for the thermal
conductivity discussed in previous section is obviously broken in the limit $T,
v_f{\rm v}_s\lesssim\gamma$ under the conditions we consider.
We note nonmonotonous superflow dependence of the thermal conductivity
upon condensate flow field under the condition $T\lesssim \gamma$, since for
$v_f{\rm v}_s\lesssim\gamma$  the thermal conductivity diminishes while
for $v_f{\rm v}_s\gg\gamma$  it increases with increasing superflow field.

\section{Conclusion}

In conclusion we have examined the possibility for nonmonotonous magnetic field
dependence of the electronic thermal conduction along the magnetic field at low
temperatures for sufficiently clean superconductors with nodes of the order
parameter on the Fermi surface. We found that the contribution from low energy
quasiparticles in the intervortex space is quite important in this respect and
specific for superconductors with nodes. The effect comes from the influence of
condensate flow field in the intervortex space on the scattering by nonmagnetic
impurities of low-energy extended quasiparticles with momentum directions near
nodes of the order parameter. The scattering is considered to be sufficiently
close to the unitarity limit. We showed that the relaxation time at low energy
is a nonmonotonous function upon the condensate flow field.

Nonmonotonous magnetic field dependence of the thermal conductivity due to this
contribution may take place for type II superconductors with large
Ginzburg-Landau parameter within the temperature interval $\beta
T_c\sqrt{\frac{ \displaystyle B_{c1}}{ \displaystyle B_{c2}}} \le T\ll T_{c}$,
where numerical factor $\beta$ may be of the order of unity. These results are
in a good qualitative agreement with recent experiments\cite{fl1,fl2} on the
heavy-fermion superconductor $UPt_3$.  We explain the anomalously strong
influence of the magnetic field on the quasiparticle scattering, observing in
this compound\cite{sud}.  We obtained as well that scaling behavior of the
thermal conductivity with a single parameter
$x=\frac{T}{T_c}\sqrt{\frac{B_{c2}}{B}}$ as well as weak low-temperature
dependence of the anisotropy ratio $\kappa_{zz}/\kappa_{yy}$ in zero field,
are valid with logarithmic accuracy within the temperature interval
$\gamma\lesssim T\ll T_{c}$  for $(1,i)$-phase of $E_{2u}$-type of pairing.
Qualitatively it is quite close to that  observed for $UPt_3$ in Ref.\
\onlinecite{fl1,fl2}.  Under the same conditions $E_{1g}$ model results in more
noticeable deviations from the scaling, and in essential temperature dependence
of the ratio $\kappa_{zz}(T,B=0)/\kappa_{yy}(T,B=0)$. New test is proposed for
discrimination between candidates for the type of pairing in $UPt_3$, based on
the dependence of $\kappa_{zz}$ upon the value of transport supercurrent
flowing in thin films or whyskers along the hexagonal axis.

\section*{Acknowledgments}

We would like to thank V.P.Mineev, who was the initiator of our work on the
subject of this article, for helpful and stimulating  discussions.  We are
grateful to the authors of Refs.\ \onlinecite{fl1,fl2} for sending us the
preprints of their works before publications. One of us (Yu.S.B.) also thanks
I.~Fomin, A.~Huxley and H.~Suderow for useful discussions. Yu.S.B. is grateful
to D\'epartement de Physique de l'ENS for financial support of his stay in
France.  This research is supported in part by the Russian Foundation for Basic
Research under grants No.~96-02-16249 and No.~97-02-17545.

\begin{figure}
\caption{The relaxation time $\tau_s$ as a function of superfluid velocity
$v_s$, directed parallel to the basal plane, in the unitary limit for various
energies $\omega/\Delta_{0}(T)= 0.01 \ (1)$, \ $0.05 \ (2)$, \ $0.2 \ (3)$, \
$1 \ (4)$, \ $5 \ (5)$ and three types of pairing:  (a) -- polar phase,
$\Delta=\Delta_0(T)\cos\theta$\, , (b) -- $(1,i)$-state of $E_{1g}$-pairing,
$|\Delta|=\Delta_0(T)|\cos\theta|\sin\theta$\, , (c) -- $(1,i)$-state of
$E_{2u}$-pairing, $|\Delta|=\Delta_0(T)|\cos\theta|\sin^2\theta$ }
\label{fig.1}
\caption{The thermal conductivity $\kappa_{zz}$ as a function of superfluid
velocity $v_s$, directed parallel to the basal plane, for various temperatures
and three types of pairing: (a) -- polar phase, (b) -- $E_{1g}$-pairing, (c) --
$E_{2u}$-pairing. The same basis functions are used as for Fig.\ 1}
\label{fig.2}
\caption{Normalized thermal conductivity $\kappa_{zz}(T,v_s)/\kappa_{zz}(T,0)$
as a function of superfluid velocity $v_s$, directed along z-axis,
for unitary scatterers, various temperatures
$T/\Delta_0(T)=0.01$\, $(1)$,\, $0.025$\,(2), \, $0.05$\,
$(3)$,\, $0.1$\, $(4)$,\, $0.2$\, $(5)$
and two types of pairing: (a) -- $E_{1g}$-pairing --
$|\Delta|=\Delta_0(T)|\cos\theta|\sin\theta$, (b) --
$E_{2u}$-pairing -- $|\Delta|=\Delta_0(T)|\cos\theta|\sin^2\theta$.}
\label{fig.3}
\end{figure}


\begin{references}
\bibitem{f} To be more careful one should say, that in Figures we plot
functions on the variable $p_fv_s/\Delta_0$ under the fixed values of
dimensionless parameters $\omega/\Delta_0$ or $T/\Delta_0$. The quantity
$\Delta_0$ should be considered self-consistently and, generally speaking,
under the condition $\Delta_0\lesssim p_fv_s $ it may essentially depend upon
$v_s$. We are concentrating, however, mostly on those values of $p_fv_s$, which
satisfy the condition $p_fv_s\ll \Delta_0$.
\end{references}
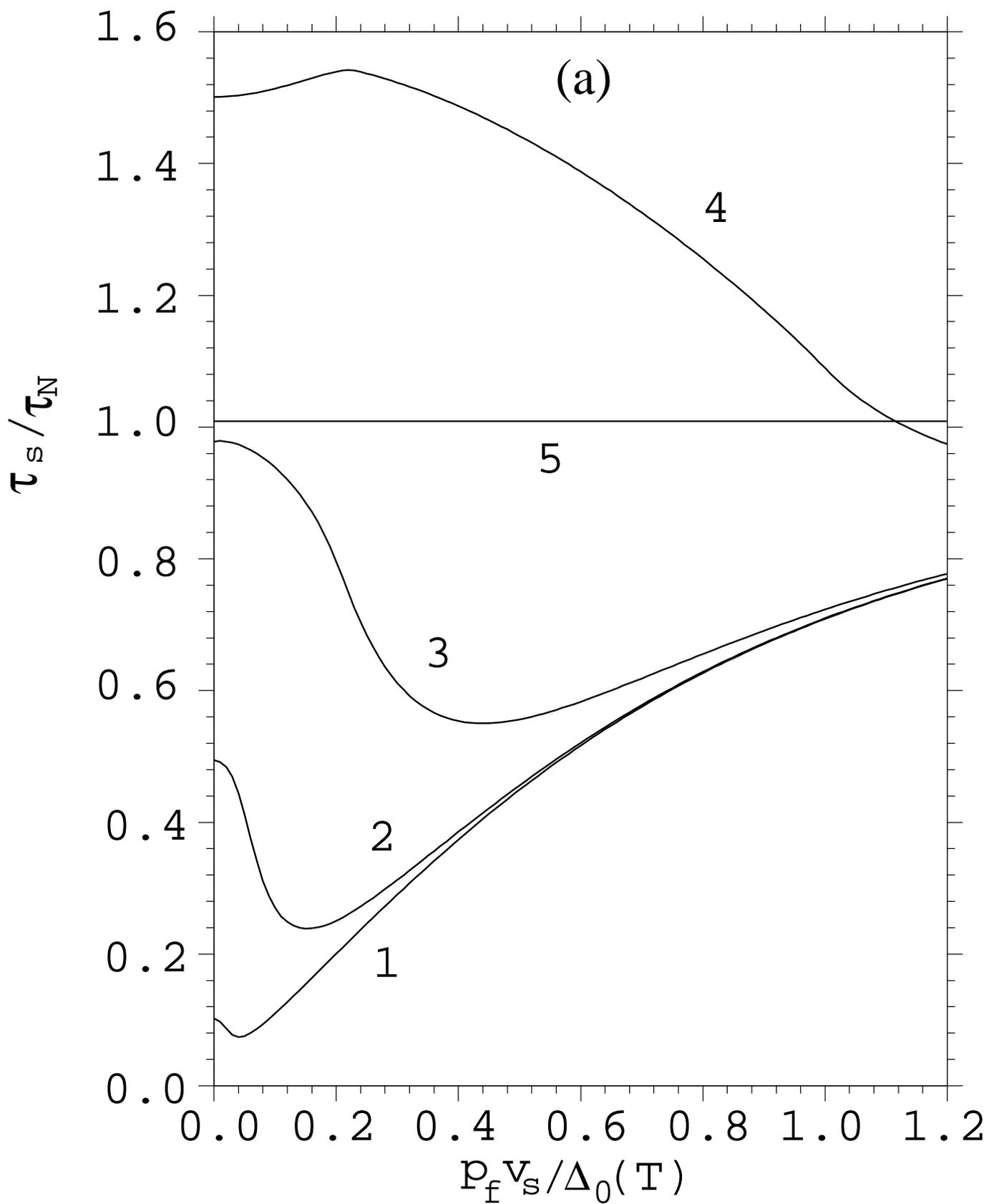
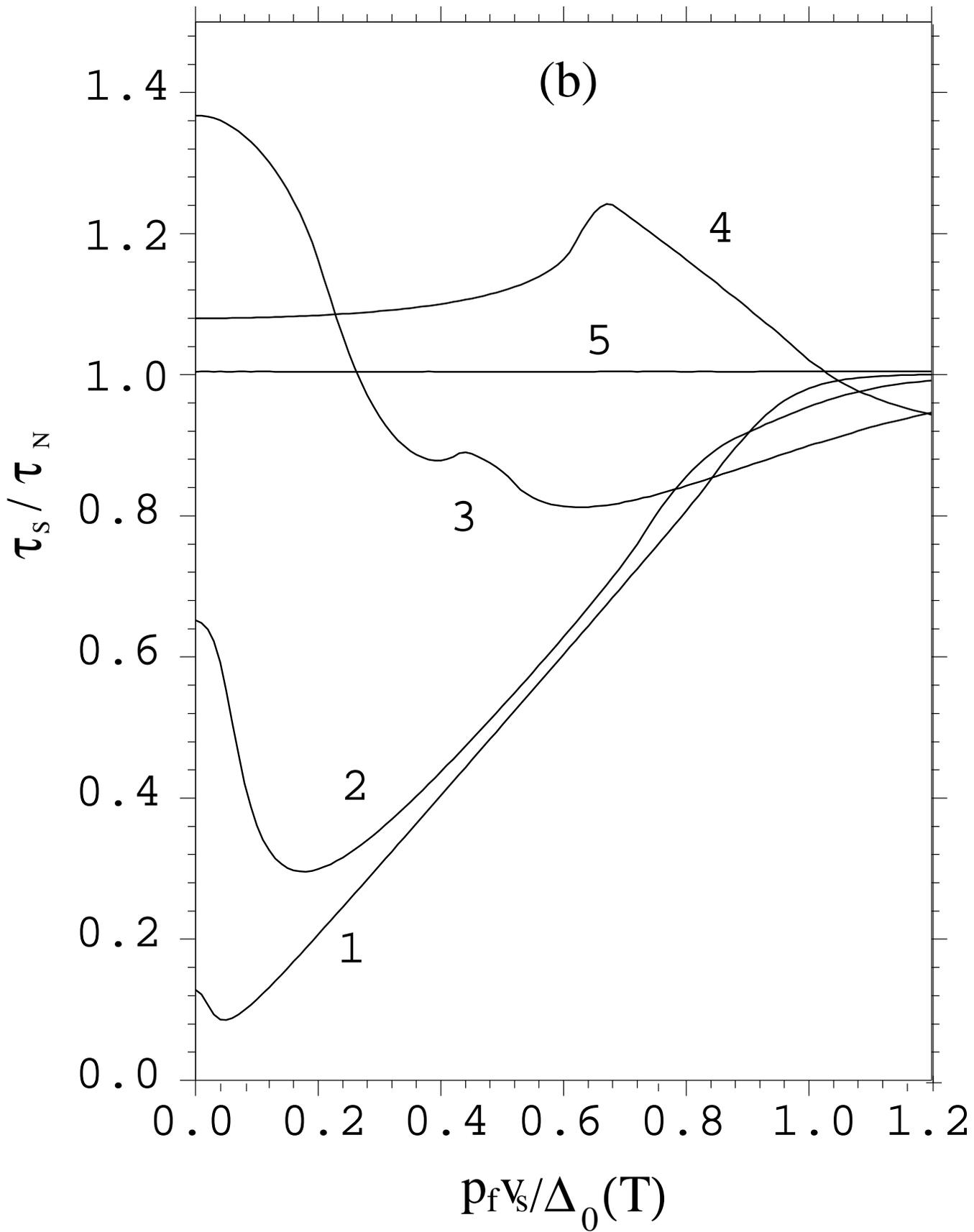
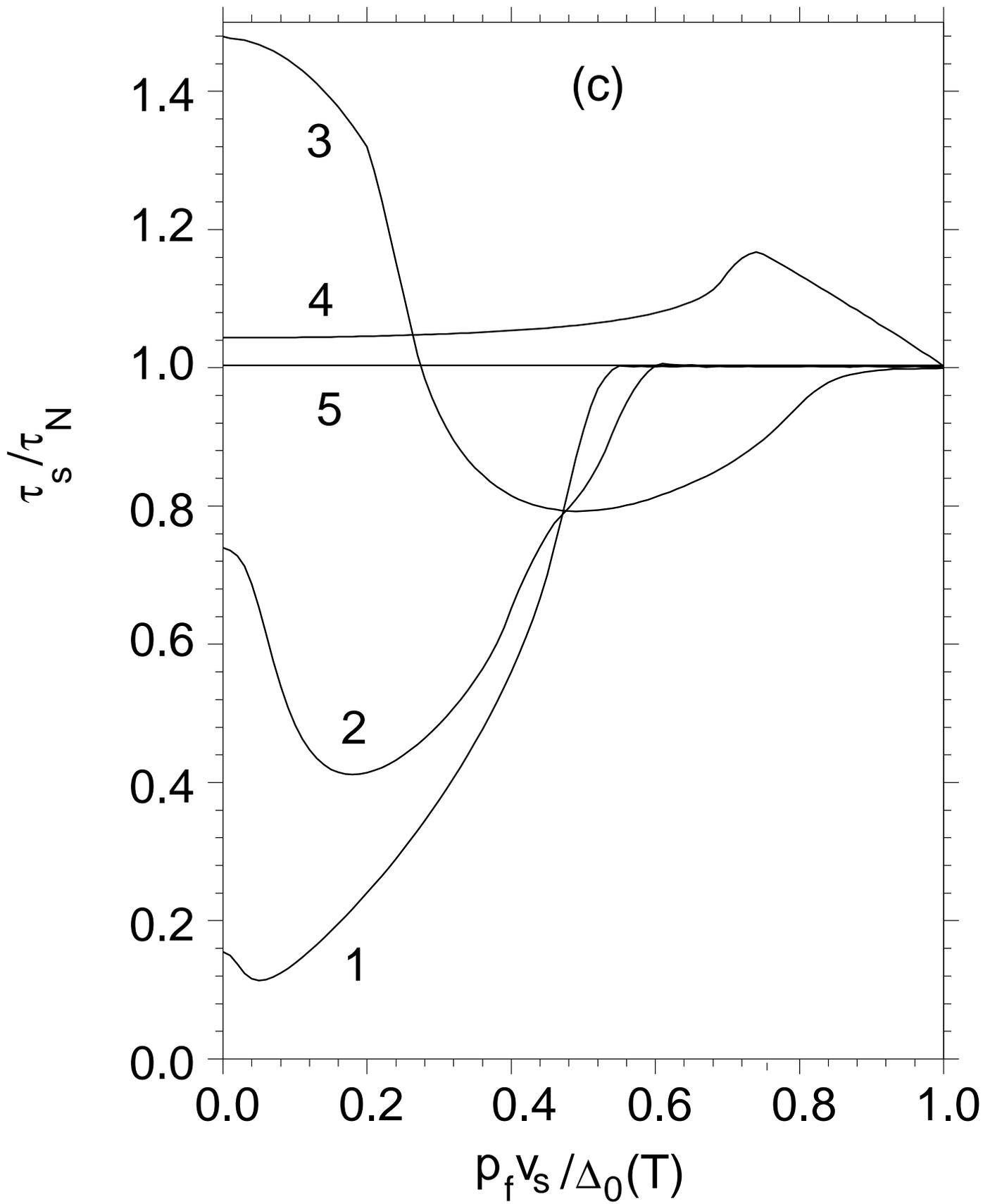
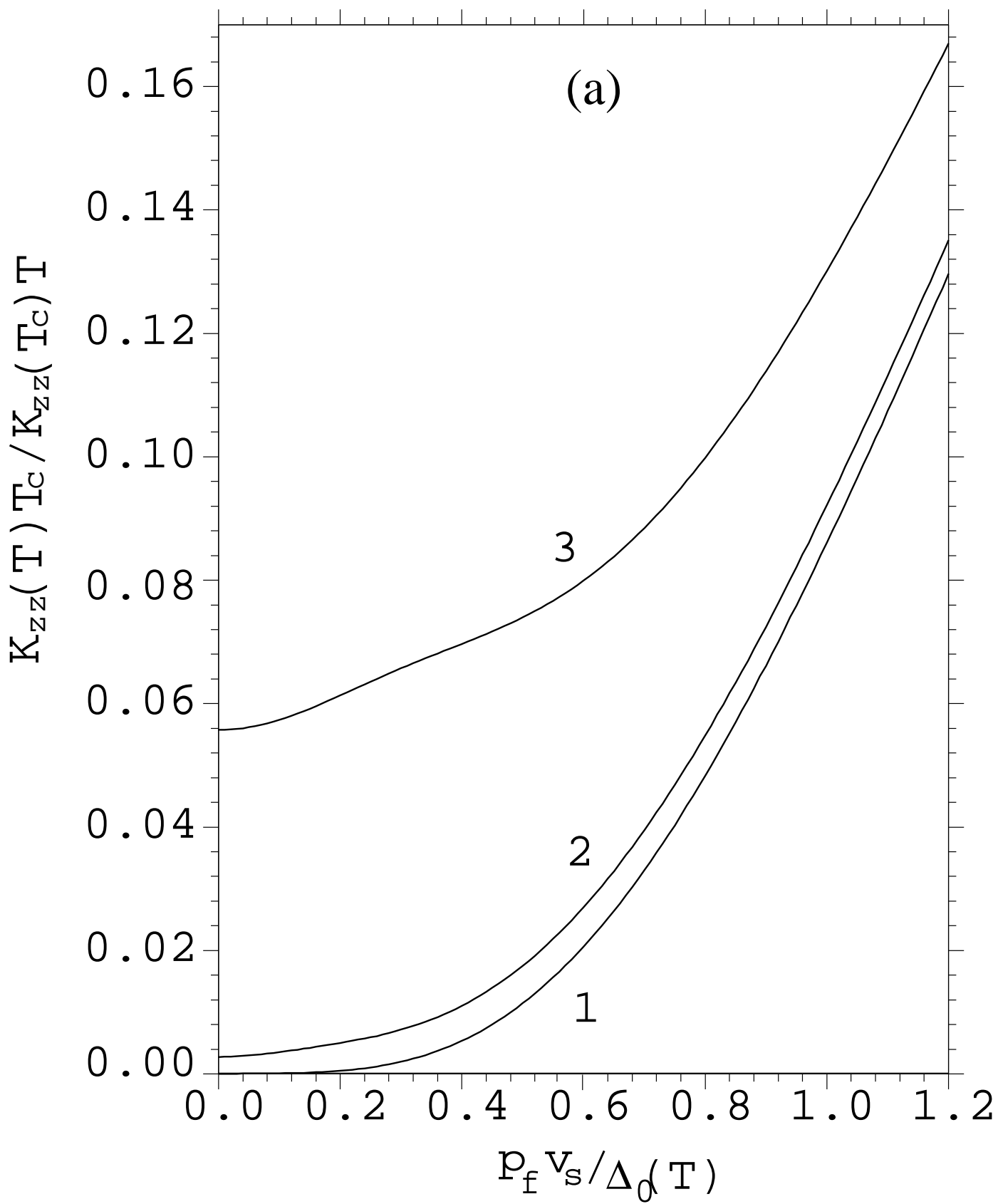
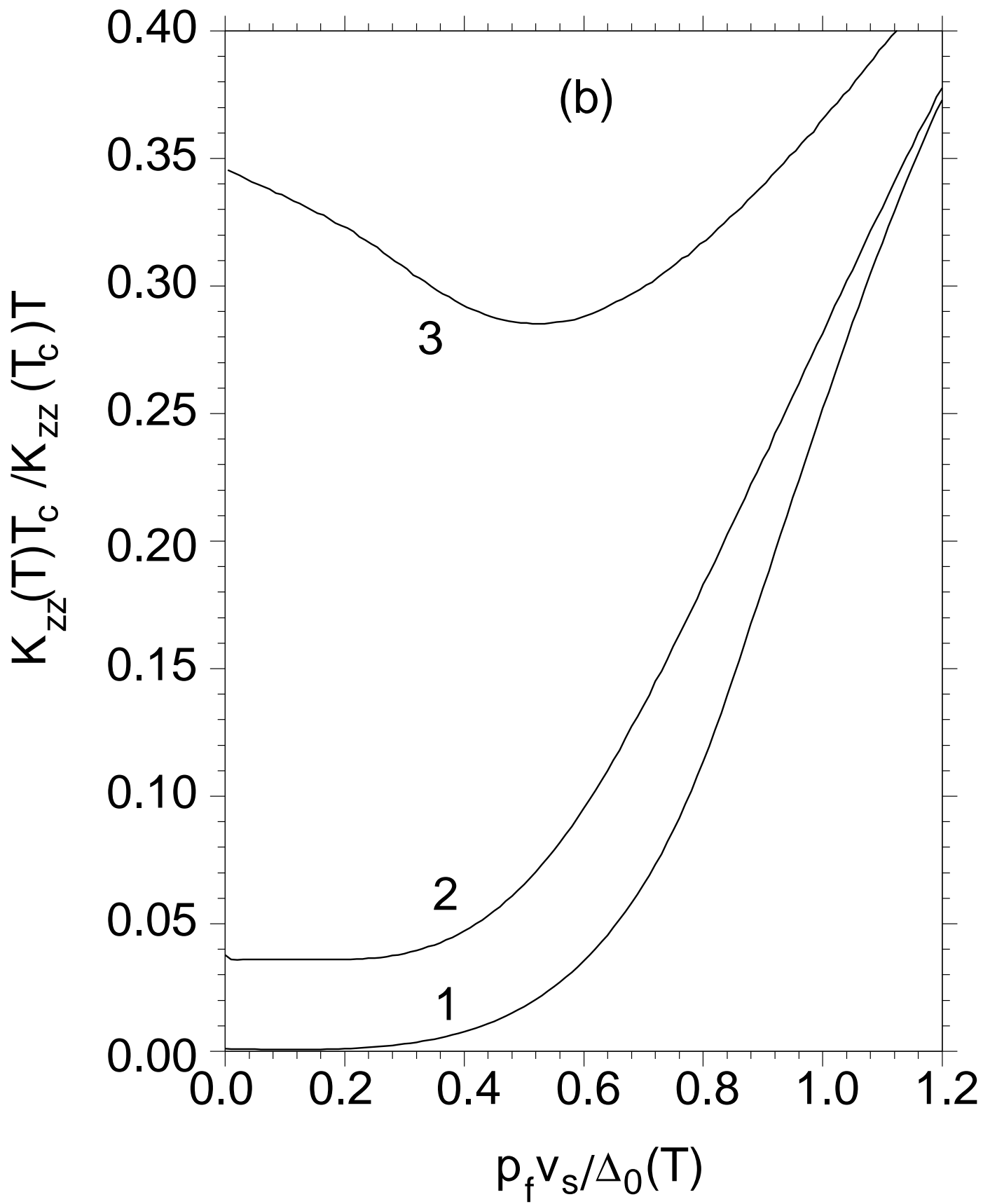
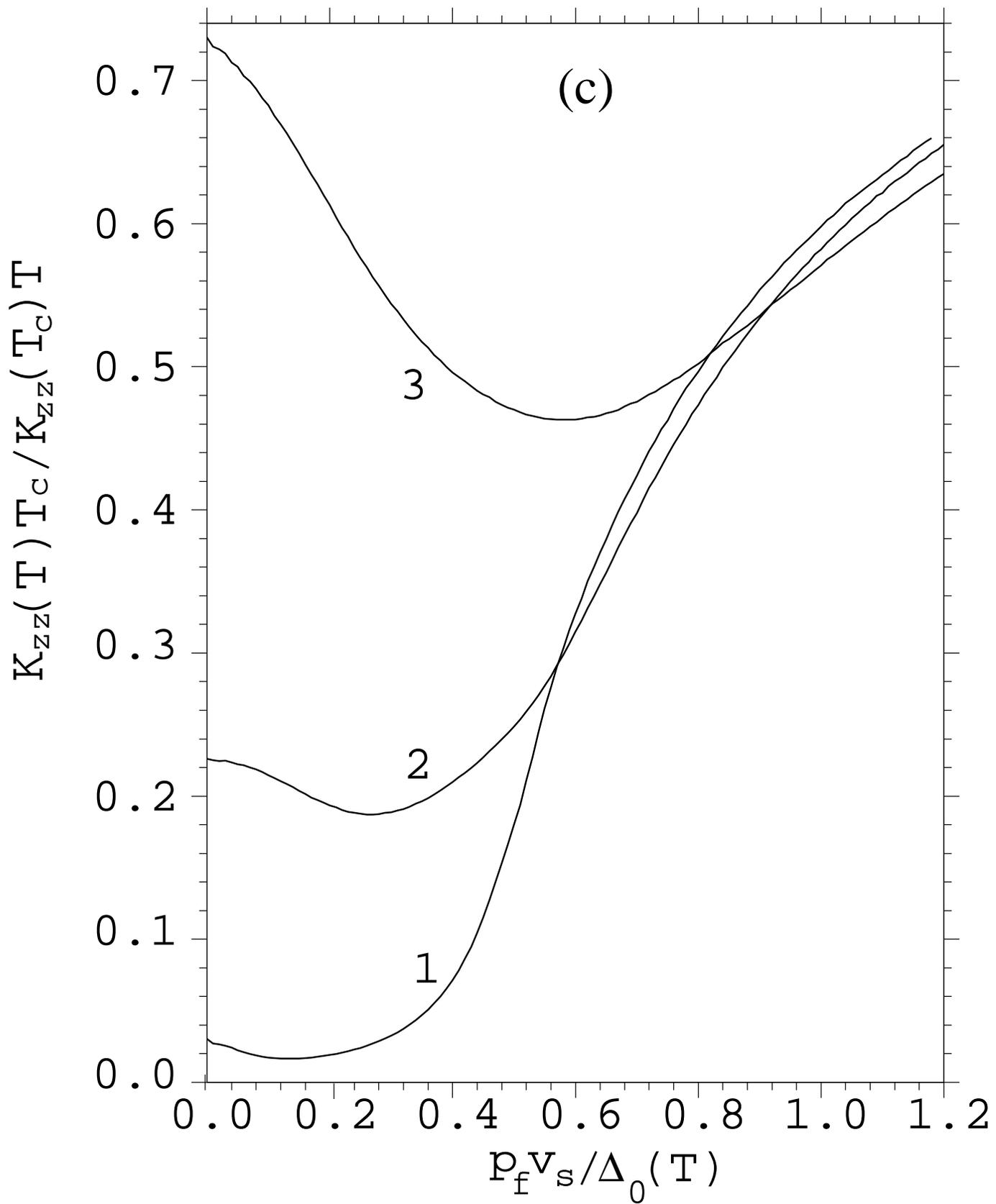
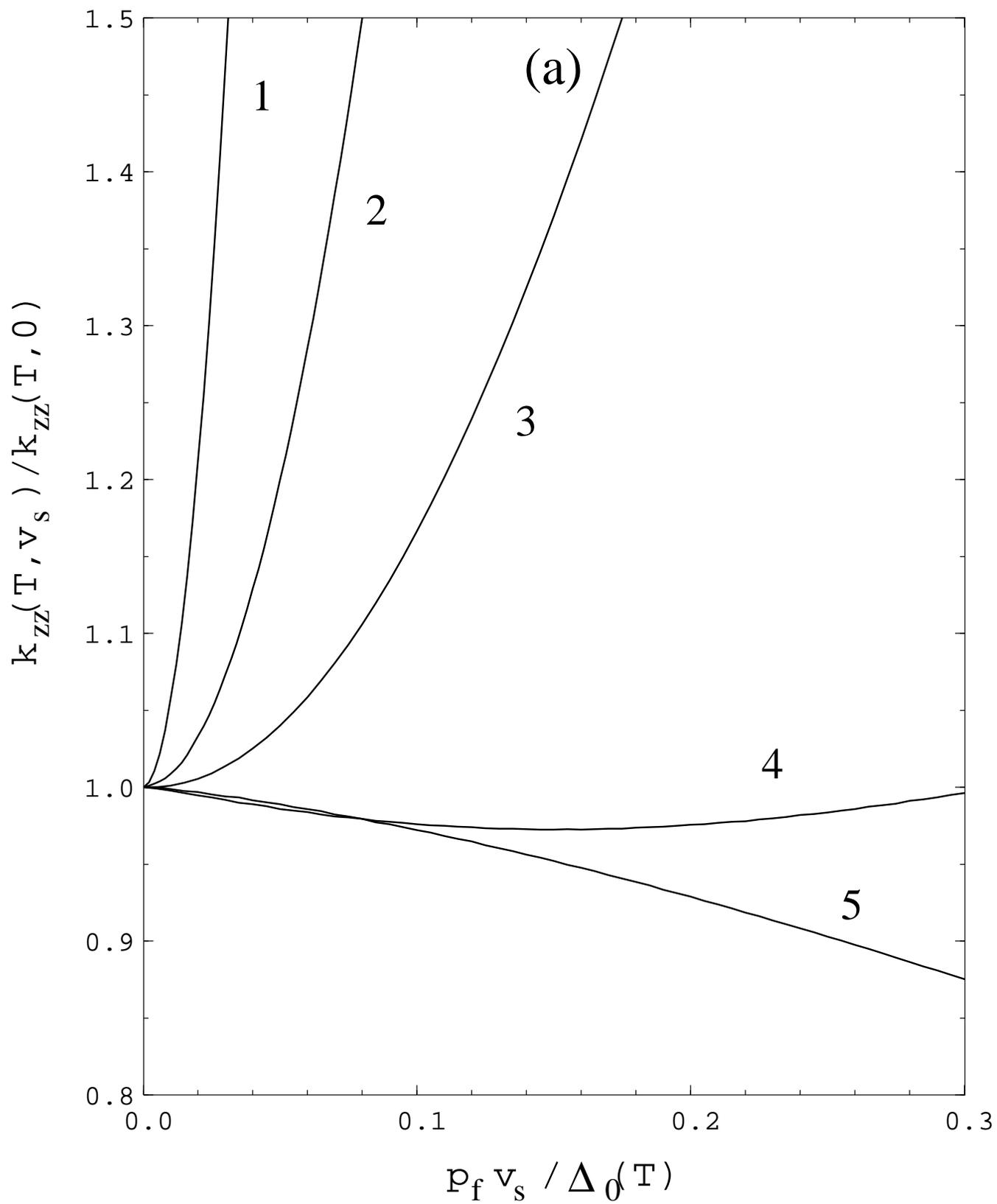
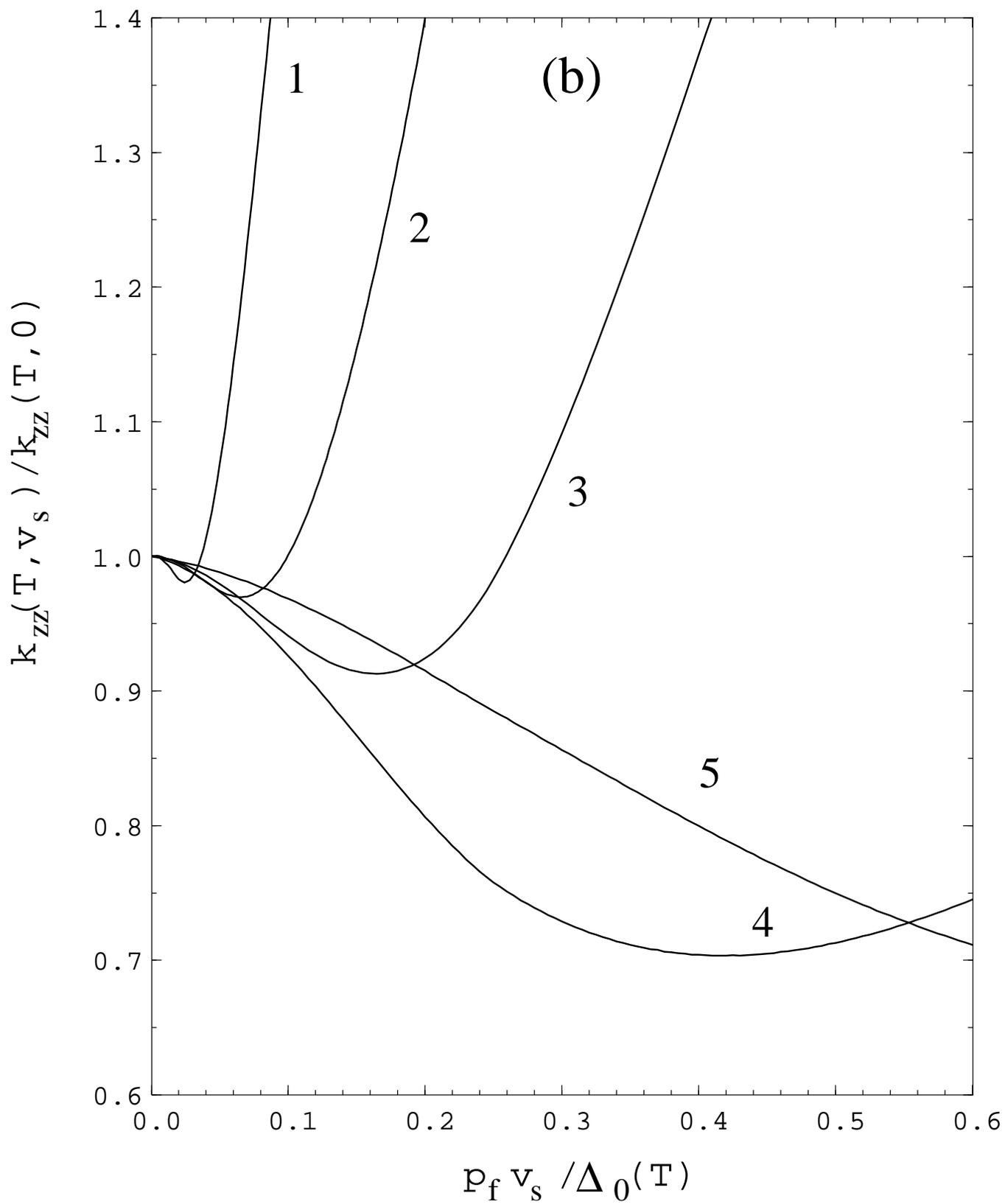
\end{document}